\documentclass[11pt]{article}

\usepackage[english]{babel}
\usepackage{graphicx}
    \usepackage{geometry}
\usepackage{setspace}
\usepackage[round, authoryear]{natbib}
\usepackage{multibib}
\newcites{supp}{References for Supplementary Materials}

\RequirePackage[OT1]{fontenc}
\RequirePackage{amsthm,amsmath,amsfonts}
\usepackage{mathtools}
\RequirePackage[colorlinks,linkcolor=blue,citecolor=blue,urlcolor=blue]{hyperref}
\usepackage{color}
\usepackage[normalem]{ulem}
\usepackage{rotating}
\usepackage{float}
\usepackage{authblk}

\usepackage{xcolor}
\newcommand{\com}[1]{\textcolor{black}{#1}}

\def \bsm {\boldsymbol}

\def \blambda {\boldsymbol{\lambda}}

\theoremstyle{plain}

\newtheorem{theorem}{Theorem}[section]

\theoremstyle{definition}

\theoremstyle{remark}

\usepackage{siunitx}

\title{Density ratio model for multiple types of survival data with empirical likelihood}
\author[1]{James Hugh McVittie}
\author[2]{Archer Gong Zhang\thanks{Corresponding author. Email address: \url{archer.zhang@glasgow.ac.uk}. \par\noindent\hspace*{1.5em} The two authors contribute equally to this work.}}
\affil[1]{Department of Mathematics and Statistics, University of Regina}
\affil[2]{School of Mathematics and Statistics, University of Glasgow}

\date{}

\begin{document}
\maketitle

\begin{abstract}
The density ratio model (DRM) is a semiparametric model that relates the distributions from multiple samples to a nonparametrically defined reference distribution via exponential tilting, with finite-dimensional parameters governing their differences in shape. When multiple types of partially observed (censored/truncated) failure time data are collected in an observational study, the DRM can be utilized to conduct a single unified analysis of the combined data. 
In this paper, we extend the methodology for censored length-biased/truncated data to the DRM framework and formulate the inference using empirical likelihood. 
We develop an EM algorithm to compute the DRM-based maximum empirical likelihood estimators of the model parameters and survival function, and assess its performance through extensive simulations under correct model specification, overspecification, and misspecification, across a range of failure-time distributions and censoring proportions.
We also illustrate the efficacy of our method by analyzing the duration of time spent from admission to discharge in a Montreal-area hospital in Canada.
The R code that implements our method is available on GitHub at \href{https://github.com/gozhang/DRM-combined-survival}{DRM-combined-survival}. 
\end{abstract}


\section{Introduction}
\label{sec:intro}

In a single observational study where multiple types of partially observed data are available or where the participant-level data from multiple independent studies are combined, it is crucial that the statistical analysis efficiently utilizes all available data \citep{riley}. For example, in a windowed sampling study design, where subjects are observed if their failure time impinges on a preset sampling window, the observed data can consist of a combination of right-censored failure times and left-truncated right-censored failure times \citep{barth, campo, shaba, welch}. In contrast, multiple data sets of the same type of failure time data can be merged into a single data set to model the failure time survival function and the factors affecting the failure time distribution \citep{mcvit, wolfs}. \com{The analysis of right-censored failure time data or left-truncated failure time data, separately, have been well studied in the literature. For a sample of right-censored failure time data, the Kaplan--Meier estimator is the nonparametric maximum likelihood estimator (NPMLE) of the failure time survival function \citep{kapla}. \cite{wang} proposed an adjusted version of the Kaplan--Meier estimator to account for the presence of left-truncation. Later, \cite{asgha} and \cite{asgha2} derived the survival function NPMLE for length-biased (i.e. uniform left-truncation distribution) right-censored data and established the estimator's asymptotic properties. The analysis of length-biased right-censored failure time data drawn from prevalent cohort studies with follow-up continues to be actively researched with recent results in diverse areas including the derivations of uniform confidence bands for hazard functions, high-dimensional variable selection, dependence measures and density ratio modelling \citep{he, rabhi, shari, zhu}.}

\com{Various} nonparametric and semiparametric approaches exist for estimating the failure time survival function using combined right-censored and left-truncated right-censored data \citep{erick, hartm, mcvit3, qin}. \com{For example, \cite{wolfs} derived the survival function NPMLE based on the Kaplan--Meier product limit estimator using a combined set of right-censored and generally left-truncated right-censored failure time data. Later, \cite{mcvit} and \cite{mcvit2} used an extended version of the EM algorithm of \cite{qin2} to derive the survival function NPMLE when right-censored and length-biased right-censored data were combined. This approach extended the work of \cite{asgha} and \cite{asgha2} to the setting where some unbiased right-censored failure times were observed in addition to the length-biased right-censored failure times. We note, however, that all the above approaches assume that the underlying failure time distributions from the unbiased right-censored cases and the biased right-censored cases are the same. Violations of this assumption can negatively affect inferences as the resulting estimators will not be targeting the correct failure time distribution. \cite{mcvit3} proposed an extension of the two-sample log-rank hypothesis test to determine whether there were any differences in the underlying failure time distributions for a sample of right-censored data when compared to a sample of left-truncated right-censored data. They applied their test to NBA and NHL datasets to determine whether certain introduced league policies altered the career lengths of the players where one group's careers began before the policy change but were still playing after the policy was introduced (i.e. left-truncated right-censored failure times) and the other group's careers began after the policy change (i.e. right-censored failure times). Surprisingly, their test revealed that certain league policy changes resulted in differences between the failure distributions of the right-censored data and the left-truncated right-censored data.}

\com{To allow for differences in the samples' failure time distributions,} we consider the density ratio model (DRM) \citep{ander}, a semiparametric approach that links the observed data distributions to a common, unspecified reference distribution via exponential tilting \com{for a combined sample of right-censored and length-biased right-censored failure time data. Specifically, we assume the underlying failure time distributions of the two samples are not necessarily equivalent and are related through the semiparametric structure of the exponentially tilted failure time distribution. That is, the right-censored sample's underlying failure time distribution is set as a reference distribution and the length-biased right-censored sample's underlying failure time distribution is exponentially tilted relative to the reference distribution.} This formulation avoids strong parametric assumptions while preserving information shared across data sources and \com{is a direct extension of the survival function NPMLE methodology of \cite{mcvit} as it does not assume the underlying failure time distributions from both samples are necessarily equivalent. Thus, the general exponential tilting model encompasses the special case where the underlying failure time distributions from the two samples are equivalent \citep{mcvit} but also when multiple samples are combined through various sources and there is doubt as to whether there are differences in the failure time distributions as in the applications of \cite{mcvit3}. Additionally, this work extends the DRM research of \citep{shen, wei, zhu, zhu2} which were focused on a single set of left-truncated right-censored failure time as well as the recent work of \cite{cai} which examined the DRM when multiple right-censored samples were combined. Other related works on the empirical likelihood (EL)-DRM framework in the literature include} numerical algorithms for estimating both the finite-dimensional parameters and the nonparametric reference distribution, as well as covariate-augmented extensions \citep{luo, huang2014joint, zhu, zhang}.  

In this paper, we develop a DRM-based procedure to estimate both the exponential tilt parameters and the unspecified reference distribution, from which the survival function estimator can be derived, when right-censored and \com{length-biased} right-censored failure time samples are analyzed jointly. We define the relevant notation for the various types of survival data for the DRM in Section~\ref{sec:method} and explain how the EL and an expectation-maximization (EM) algorithm-based procedure can be used to nonparametrically estimate the unknown reference distribution function and unknown parameters in the DRM. In Section~\ref{sec:simulation}, we perform an extensive simulation study to highlight the efficacy and accuracy of the DRM-based survival function estimator when combining multiple data sources and in Section~\ref{sec:application}, we utilize the DRM to model the distributions for the time spent from admission to discharge in a Montreal-area hospital. Finally, Section~\ref{sec:discussion} elucidates the connection between arbitrary left-truncated failure time data and the DRM, and provides some discussions and directions for future research work. 
For reproducibility, the {\tt R} \citep{R} code for the proposed DRM-based survival estimation method, together with implementations of the competing estimators considered in this paper and an example script demonstrating the simulation procedure, is publicly available on GitHub at \href{https://github.com/gozhang/DRM-combined-survival}{DRM-combined-survival}.

\section{Methodology}
\label{sec:method}

In most nonparametric estimation problems, it is assumed that the observed random variables of the sample are drawn from the density/cumulative distribution function of interest. In these settings, the primary goal is to derive a nonparametric estimator of the density function or cumulative distribution function with minimal assumptions on its shape. In some cases where multiple samples are available from different populations, the observed random variables may not be drawn from the density function of interest and may instead be drawn from an exponentially tilted version of the target density function \citep{luo}. Let $F_0(\cdot)$ denote a reference distribution function, let $h(\cdot)$ denote some prespecified basis function (possibly vector-valued), and let $\theta$ denote an unknown parameter (possibly a vector). 
For expository simplicity, we assume without loss of generality that both $h(x)$ and $\theta$ are scalars. 
The results and conclusions throughout this paper remain valid when they are vector-valued.
The DRM relates the density function of the observed data to the reference density function via the following expression:
\begin{align}
\mathrm{d}F(x; \theta) = \exp\{\alpha + \theta h(x)\} \mathrm{d}F_0(x),
\label{DRM}
\end{align}
where $\alpha = -\log\left(\int \exp\{\theta h(x)\} \mathrm{d}F_0(x) \right)$ is the normalizing constant. \com{When multiple sources of data are combined,} under the DRM, \com{we assume some subset of data is drawn from $F(x; \theta)$, and the remaining subset is drawn from $F_0(x)$.} \com{Our goal is to }estimate the unknown parameter $\theta$ and the unspecified distribution function $F_0(x)$ using the combined data. Throughout the remainder of this paper, when multiple sources of data are combined, we will specify whether the samples are drawn from the reference distribution $F_0(x)$ or from the exponentially tilted distribution $F(x; \theta)$.
\com{We also remark that the choice of the basis function $h(\cdot)$ is part of the DRM specification. In applications, it is typically guided by scientific knowledge or exploratory distributional features. For example, one may use basis functions such as $(x,x^2)^\top$ when the samples resemble normal-type distributions, or $(x,\log (x))^\top$ when they resemble gamma-type distributions, so that these common parametric families are contained in the corresponding DRM. Another common strategy is to start with a richer collection of elementary functions and then select a parsimonious model using a selection criteria such as Akaike or Bayesian information criterion \citep{fokianos2007density}. More recently, data-adaptive basis construction, such as the functional principal component analysis based approach of \citet{zhang}, has also been proposed when sufficiently many samples or data sources are available. In this paper, we take $h(\cdot)$ to be user-specified and examine several common choices in the numerical studies and application.}

Prior to discussing the setting when multiple sources of data are available for analysis, we consider analyses using a single set of partially observed failure time data. When a single source of data is available, it can either be assumed that the data are drawn from the reference density function or from its exponentially tilted version. The latter case is typically never assumed as it requires that additional population information must be known to ensure the DRM components are identifiable, and in that case the model essentially collapses to the usual nonparametric setting \citep{zhang2}. Thus, when only one source of data is available, in most applied settings, there is no reason to assume the sampled data are not drawn from the reference density unless some additional information regarding the population distribution structure is known. In the one-sample setting, we will restrict our analyses to the case where the data are drawn from the reference density and highlight how the empirical likelihood can be used to nonparametrically estimate the survival function.

We first introduce the notation for the considered survival data. Let $T$ denote the failure time of interest measured from an initiating event to a failure event. Depending on the study design, the failure time may be partially observed due to the features of censoring and/or truncation. We restrict our analysis to right-censored and left-truncated right-censored failure time data. Let $C$ denote the right-censoring random variable and let $A$ denote the left-truncation random variable. When the underlying left-truncation distribution is assumed to be uniformly distributed, we will refer to the set of the left-truncated right-censored failure time data as length-biased right-censored failure time data \citep{asgha, asgha2}. 

\subsection{Single-Sample Methods}

For a single sample of partially observed failure time data, the assumptions on the censoring/truncation mechanism have a direct impact on the resulting estimation procedures \citep{aalen}. Although the single-sample approaches have already been thoroughly discussed in the statistical literature, we nonetheless highlight how the empirical likelihood framework \citep{owen} and EM algorithm method \citep{demps} are connected to commonly used estimation procedures. 
\com{We note that, while these methods can be described directly through nonparametric likelihood, we present them from an empirical likelihood perspective to provide a natural transition to the DRM-based empirical likelihood formulation developed in later sections for combining multiple samples.}

\subsubsection{Independent and Identically Distributed Data}

Let $X_1, \ldots, X_n$ denote an observed sample of independent and identically distributed (i.i.d.) failure time data. The empirical likelihood of $F_0$, based on the observed data, is given by:
$$L(F_0) = \prod_{i=1}^{n} \mathrm{d}F_0(x_i).$$
Numerous techniques in the nonparametric density function estimation literature can be applied to derive an estimator for $\mathrm{d}F_0$ or $F_0$. A standard approach for estimating the function $F_0$ is through an empirical likelihood approach, which assumes $\mathrm{d}F_0$ takes non-zero probability values, $p_1, \ldots, p_k$, on the unique time points $t_1<\cdots<t_k$, respectively. Estimators for the probability masses $p_1, \ldots, p_k$ are then determined by maximizing the empirical log-likelihood $\log L(F_0)$ subject to the constraint $\sum_{j=1}^k p_j=1$. 
Such a constrained optimization problem can be solved by the method of Lagrange multipliers, with the Lagrangian function given by:
$$\mathcal{L}(p_1, \ldots, p_k, \lambda) = \sum_{j=1}^{k} r_j\log(p_j) + \lambda \{1 - \sum_{j=1}^{k} p_j\},$$
where $r_j = \sum_{i=1}^{n} 1(X_i = t_j)$ and $\lambda$ is the Lagrange multiplier corresponding to the constraint. Constrained maximization of the above empirical likelihood has the solution $\hat p_j=r_j/n$, which yields the standard empirical cumulative distribution function given by $\hat{F}_{0}(x) = n^{-1} \sum_{i=1}^{n} 1(X_i \leq x)$ as the estimator of $F_0$.

\subsubsection{Right-Censored Data}

Consider the case where the observed failure time data are partially observed by a random right-censoring variable. This situation occurs frequently in medical/clinical studies where the failure event is unobserved due to some other factors or loss during follow-up \citep{mcvit}. Let $X = \min(T, C)$ denote a right-censored failure time and let $\delta = 1_{\{T < C\}}$ denote the indicator variable as to whether the failure time is right-censored. The sample consists of the pairs of observations $(X_i, \delta_i)$ for $i = 1, \ldots, n$. We assume the right-censoring random variables $C$ are independent of $T$ (i.e. the random censoring assumption) and that $C$ non-informatively right-censors $T$ (i.e. the distribution of $C$ yields no information on the distribution of $T$) \citep{aalen}. The observed data likelihood function is given by:
$$L(F_0) \propto \prod_{i=1}^{n} \left[ \mathrm{d}F_0(x_i) \right]^{\delta_i} \left[1 - F_0(x_i)\right]^{1-\delta_i}.$$
As in the i.i.d. setting, since the censoring mechanism is non-informative, we assume $\mathrm{d}F_0$ places non-zero probability masses $p_1, \ldots, p_k$ on the unique observed failure times $t_1 < \cdots < t_k$, respectively, to form the empirical likelihood function:   
\begin{equation}
L(p_1, \ldots, p_k) = \prod_{j=1}^{k}  p_j^{r_j} \left[1 - \com{\sum_{l: t_l \leq t_j} p_l}\right]^{\xi_j},
\label{EL_RC}
\end{equation}
with $r_j = \sum_{i=1}^{n} I(X_i = t_j, \delta_i = 1)$ and $\xi_j = \sum_{i=1}^{n} I(X_i = t_j, \delta_i = 0)$. \com{We note that the nonparametric maximum likelihood estimator (NPMLE) of the failure time survival function, determined from the above empirical likelihood, corresponds to the Kaplan--Meier estimator \citep{kapla}. As our derivation of a survival function estimator using multiple samples (described in later sections) is based on the EM algorithm, } to maximize the above empirical likelihood function while respecting the constraint that $\sum_{j=1}^{k} p_j = 1$, we \com{present the details of how the EM algorithm can be utilized in this setting} \citep{demps}.

\com{Let the union of the observed failure times $X_i$ with $\delta_i = 1$, for $i = 1, \ldots, n$ and the unobserved failure times, $T_i$, drawn from $F_0(\cdot)$, corresponding to the right-censored lifetimes $X_i$ with $\delta_i = 0$ for $i = 1, \ldots, n$, denote the complete data.} Let $\hat p_j^{(m)}$ denote the probability mass for $T_j$ evaluated at the $m$th iteration of the EM algorithm. The expected complete-data log-likelihood, conditional on the observed data $\mathcal{O}$ and the $m$th iterates of the parameter estimates $\hat{p}_1^{(m)}, \ldots, \hat{p}_k^{(m)}$ is given by
$$\ell_c(p_1, \ldots, p_k) = \sum_{j=1}^{k} (r_j + w_j^{(m)}) \log(p_j),$$
with 
\[
w_j^{(m)} = \mathbb{E}[I(X_i = t_j, \delta_i = 0) | \mathcal{O}] = \sum_{i=1}^{n} (1 - \delta_i) \frac{\hat{p}_j^{(m)} I(X_i \leq t_j)}{\sum_{j'=1}^{k} \hat{p}^{(m)}_{j'} I(X_i \leq t_{j'})},
\]

Since the probability masses must satisfy $\sum_{j=1}^{k} p_j = 1$, we still use the Lagrange multiplier method. 
Let $\lambda$ denote the Lagrange multiplier for this constraint. 
The Lagrangian is given by 
$$\mathcal{L}_c(p_1, \ldots, p_k, \lambda) = \sum_{j=1}^{k} (r_j + w_j^{(m)}) \log(p_j) + \lambda\{1 - \sum_{j=1}^{k} p_j\}.$$ 
Thus, the EM algorithm proceeds with initial estimates $\hat{p}_1^{(0)}$, \ldots, $\hat{p}_k^{(0)}$, computing $w_j^{(0)}$ using these estimates, and then solving the above Lagrangian system with respect to the unknown parameters $p_1, \ldots, p_k$. It can be shown that the solution of $p_j$ that maximizes the above log-likelihood under the constraint satisfies the iterative update equation:
$$\hat{p}_j^{(m+1)} = \frac{r_j + w_j^{(m)}}{n}.$$
The resulting empirical likelihood estimator of $F_0$ based on the probability masses $\hat p_1^{(m)}, \ldots, \hat p_k^{(m)}$ coincides with the nonparametric estimator of $F_0$ studied by \cite{efron}, and therefore converges to the standard Kaplan--Meier product-limit estimator as the number of steps $m$ tends to infinity \citep{kapla}.

\subsubsection{Left-Truncated Right-Censored Data}

In a cross-sectional medical study, subjects are typically screened through a test or survey, and only those individuals who were prevalent for the condition of interest are followed forward. This sampling design produces left-truncated right-censored failure time data, as not all subjects are recruited into the study. A set of left-truncated right-censored failure time data consists of the triples, $(A_i, X_i=\min(T_i, A_i + C_i), \delta_i)$ such that $T_i > A_i$ for $i = 1, \ldots, n$. \com{\cite{wang} derived the full likelihood function for a set of left-truncated right-censored failure time data and factored it into the product of two likelihoods $L_1(F_0)$ and $L_2(F_0)$ where 
$$L_1(F_0) \propto \prod_{i=1}^{n} \frac{[\mathrm{d}F_0(x_i)]^{\delta_i} [1 - F_0(x_i)]^{1 - \delta_i}}{1 - F_0(a_i)}.$$
Interestingly, Wang showed that the NPMLE of $F_0$ based on $L_1(F_0)$ is equal to the NPMLE of $F_0$ based on the full likelihood function $L_1(F_0)L_2(F_0)$. The NPMLE of $1 - F_0(\cdot)$ takes the same form as the Kaplan--Meier estimator where the risk sets account for both the failure and left-truncation times (i.e. the risk set at time $t$ is given by $R(t) = \sum_{i=1}^{n} I(A_i < t < X_i)$).} We note that in the above setting \com{of \cite{wang}}, the truncation distribution is assumed to be unknown. 

In a variety of applications in medicine, biology, and engineering, a common assumption is that the left-truncation times are uniformly distributed. This specific type of left-truncation distribution occurs when the event process for the initiating events of the failure times is assumed to be a stationary Poisson point process \citep{asgha}. Under this assumption, the observed failure time data are commonly referred to as length-biased right-censored (LBRC) failure time data. Unlike the above conditional likelihood for the general truncation setting, the observed likelihood for LBRC data can be expressed as:
$$L(F_0) \propto \prod_{i=1}^{n} \frac{\left[\mathrm{d}F_0(x_i)\right]^{\delta_i}[1-F_0(x_i)]^{1-\delta_i}}{\mu(F_0)}$$
where $\mu(F_0) = \mathbb{E}(T)$. \cite{asgha} and \cite{asgha2} derived the NPMLE of the survival function through a two-step procedure by first deriving the NPMLE of the length-biased failure time survival function and then using it as a plug-in estimator for the unbiased failure time survival function. Later, it was shown by \cite{qin2} that the unbiased failure time survival function can be nonparametrically estimated directly through an EM algorithm. We present their EM algorithm procedure here and highlight the role of the empirical likelihood. 

Under the assumption of a stationary Poisson process for the initial events of the failure times, it can be shown that the right-censoring mechanism is informative for the failure-time distribution. Thus, we let $t_1 < \cdots < t_k$ denote the ordered unique failure/censoring times with respective probability masses $p_1, \ldots, p_k$ such that $\sum_{j=1}^{k} p_j = 1$ (i.e. we allow positive probability mass on both the observed failure and right-censoring times). Let $r_j = \sum_{i=1}^{n} I(x_i = t_j, \delta_i = 1)$ and $\xi_j = \sum_{i=1}^{n} I(X_i = t_j, \delta_i = 0)$ denote the number of failure times and censoring times at the time point $t_j$, respectively. We note that due to the presence of the uniform left-truncation distribution, there is a random number, $M$, of left-truncated failure times $T^*$, that were unobserved as $T^* < \com{A}$. That is, their truncation times were larger than the corresponding failure times. This implies that the complete data consists of the observed failure times, the failure times corresponding to the right-censoring times, as well as the failure times corresponding to the unobserved left-truncated failure times. Therefore, the complete-data empirical log-likelihood function can be expressed as:
\[
\ell_c(p_1, \ldots, p_k) = \sum_{j=1}^{k} \left\{\sum_{i=1}^{n} I(T_i = t_j) + \sum_{i=1}^{M} I(T^*_i = t_j) \right\} \log (p_j).
\]
\com{In} evaluating the conditional expectation of the first two terms inside the curly brackets, the first expectation will simplify to the same expectation as for the right-censored failure time data case. As the second summation consists of a random number of i.i.d. indicator functions, we apply Wald's equation to obtain:
\[
\mathbb{E}\left[\sum_{i=1}^{M} I(T_i^* = t_j) | \mathcal{O}\right] = \mathbb{E}(M | \mathcal{O}) \mathbb{E}(I(T^* = t_j) | \mathcal{O}).
\]
Since $M$ corresponds to the number of truncated failure times with $T^* < \com{A}$, then $M$ is negative-binomially distributed with parameters $n$ and $\pi = \mathbb{P}(T < A)$. Under the assumption that the underlying left-truncation times are uniformly distributed \com{over $(0, \tau)$}, it follows that $\pi = \mathbb{E}(T)/\com{\tau}$. \com{Based on the arguments of \cite{qin2}, the length-biased failure times with discretized probability masses $\mathrm{d}F(t_i) = p_i$, $i = 1, \ldots, k$ can be generated according to $T \sim F$ and $A \sim U(0, t_k)$ where $\hat{\tau} = t_k$ subject to $T > A$. In general, for a failure time distribution defined over $(0, \tau_0)$, assuming such an upper bound ($\tau_0$) exists, the truncation parameter must satisfy $\tau > \tau_0$ whereby $t_k$ is a consistent estimator for $\tau_0$. \cite{qin2} prove that not only is the estimator consistent but it has rate faster than $\sqrt{n}$. Using the above estimator for $\tau$, an estimator for $\pi$ is given by $\hat{\pi} = \sum_{i=1}^{k} t_ip_i / t_k$.} Thus, the conditional expectation of $M$ is given by $\mathbb{E}(M|\mathcal{O}) = n(1-\pi)/\pi$, where the expectation of the left-truncated failure time indicator function is given by $\mathbb{E}(I(T^* = t_j) | \mathcal{O}) = p_j (1 - t_j/t_k)/(1 - \pi)$. Therefore, the conditional expectation of the complete-data log-likelihood yields the following expected conditional log-likelihood:
\[
\mathbb{E}(\ell_c(p_1, \ldots, p_k) | \mathcal{O}) = \sum_{j=1}^{k} \left[ r_j + \sum_{i=1}^{n} (1 - \delta_i) \frac{p_j I(X_i \leq t_j)}{\sum_{j'=1}^{k} p_{j'} I(X_i \leq t_{j'})} + \frac{n p_j(1 - \frac{t_j}{t_k})}{\hat\pi}\right] 
\log(p_j).
\]

As in the case for right-censored data, we use the method of Lagrange multipliers to solve the constrained maximization of the above log-likelihood.
With $\lambda$ denoting the Lagrange multiplier for the constraint $\sum_{j=1}^{k} p_j=1$, the Lagrangian is
\[
\mathcal{L}_c(p_1, \ldots, p_k, \lambda) = \sum_{j=1}^{k} \left[ r_j + \sum_{i=1}^{n} (1 - \delta_i) \frac{p_j I(X_i \leq t_j)}{\sum_{j'=1}^{k} p_{j'} I(X_i \leq t_{j'})} + \frac{n p_j(1 - \frac{t_j}{t_k})}{\hat\pi}\right] 
\log(p_j) + \lambda\{1 - \sum_{j=1}^{k} p_j\}.
\]
Setting initial estimates for the probability masses given by $\hat{p}_1^{(0)}, \ldots, \hat{p}_k^{(0)}$ and then solving the above constrained maximization problem, the unknown model probability masses $p_1, \ldots, p_k$ can be estimated through an iterative updating procedure. For notational convenience, we let
\[
w_{j}^{(m)} 
= r_j + \sum_{i=1}^{n} (1 - \delta_i) \frac{\hat{p}_j^{(m)} I(X_i \leq t_j)}{\sum_{j'=1}^{k} \hat{p}_{j'}^{(m)} I(X_i \leq t_{j'})} + \frac{n \hat{p}_j^{(m)}(1 - \frac{t_j}{t_k})}{\hat{\pi}^{(m)}},
\]
where $\hat{\pi}^{(m)} = \sum_{j=1}^{k} \hat{p}_j^{(m)} t_j/t_k$. 
At the $(m+1)$th iteration of the algorithm, the updated parameters $\hat{p}_j^{(m+1)}$ satisfy $\partial\mathbb{E}(\ell_c(p_1, \ldots, p_k)|\mathcal{O})/\partial p_j=0$, where after some algebra, it can be shown that $\hat{p}_j^{(m+1)} = \hat{\pi}^{(m)}w_{j}^{(m)}/n$. \cite{qin2} showed that the corresponding survival function based on the probability masses $\hat p_j^{(m)}$ as the number of steps $m$ tends to infinity corresponds exactly to the NPMLE proposed by \cite{asgha}.

\subsection{Two-Sample Methods}

In the two-sample setting under the DRM framework, it must first be assumed that either (i) both samples are drawn directly from the same reference density, (ii) only one of the samples is drawn from the reference density and the other sample is drawn from an exponentially tilted version of it, or (iii) both samples are drawn from exponentially tilted versions of a common unspecified reference density. As case (i) reduces to a purely nonparametric estimation problem studied in \citep{mcvit2}, and case (iii) requires a priori population information to impose an additional constraint to ensure model identifiability which essentially collapses to case (ii), we will not discuss either case further. Additional remarks on case (iii) are provided in Section~\ref{sec:discussion}. 
Focusing on case (ii), where partially observed unbiased and biased failure time data are combined, we remark that interpretability of the DRM in \eqref{DRM} is improved by taking the unbiased failure time density as the reference density. 
Specifically, if $F_0$ denotes the distribution of the unbiased failure times, then the exponentially tilted failure time distribution is modelled via the DRM as 
\[
\mathrm{d}F_{\text{exptilt}}(x; \theta) = \exp\{\alpha + \theta h(x)\} \mathrm{d}F_0(x),
\]
in which the parameter $\theta$ quantifies the tilt from the unbiased distribution to the exponentially tilted distribution.
By contrast, if the biased observed data distribution were taken as the reference $F_0$, then $\theta$ would quantify the tilt of the unbiased distribution relative to the biased one. 
This is less interpretable because our target is the unbiased distribution and its survival function, even though the aforementioned two parameterizations are statistically equivalent and the DRM is symmetric under swapping the reference and tilted distributions.
While some discussion of such a comparison was conducted in the context of the proportional hazards model by \cite{wang}, we will instead consider the simpler case where the reference density corresponds to the unbiased failure-time distribution. 

Suppose a set of right-censored failure time data is combined with a set of length-biased right-censored failure time data. \com{We will assume the underlying failure time distribution of the} right-censored failure times \com{corresponds to the reference distribution} and the \com{underlying failure time distribution of the} length-biased right-censored failure times \com{corresponds to the} exponentially tilted version of the reference density via the DRM. We will use the notation described in the preceding section for the single-sample setting; however, we will distinguish the failure/censoring times in the length-biased sample from those in the unbiased sample using prime notation (i.e. $T_i'$ is a length-biased failure time where $T_i$ is an unbiased failure time). 
Under the DRM assumption, the observed data likelihood is:
\begin{equation}
L(\alpha, \theta, F_0) = \left \{ \prod_{i=1}^{n_{\text{RC}}} [\mathrm{d}F_0(x_i)]^{\delta_i} [1 - F_0(x_i)]^{1-\delta_i} \right \}
\left \{ \prod_{i=1}^{n_{\text{LBRC}}} \frac{[\exp\{\alpha + \theta h(x_i')\}\mathrm{d}F_0(x_i')]^{\delta_i'} [1 - F_{\text{exptilt}}(x_i'; \alpha, \theta, F_0)]^{1-\delta_i'}}{\mu(\alpha, \theta, F_0)} \right \},
\label{2samplike}
\end{equation}
where $n_{\text{RC}}$ and $n_{\text{LBRC}}$ denote the respective sample sizes of the right-censored and the length-biased right-censored data samples \com{and $F_{\text{exptilt}}(x; \alpha, \theta, F_0) = \int_{0}^{x} \exp\left\{\alpha + \theta h(u)\right\}\mathrm{d}F_0(u)$}. Let $t_1 < \cdots < t_k$ denote the combination of the unique, ordered failure times from the right-censored data set and the unique, ordered failure/censoring times from the length-biased right-censored data set with (reference) probability masses $p_1, \ldots, p_k$, respectively, with $\sum_{j=1}^{k} p_j = 1$. Let $q_j = p_j\exp\{\alpha + \theta h(t_j)\}$ for $j = 1, \ldots, k$ denote the probability masses for times drawn from the exponentially tilted distribution (i.e. the failure-time distribution $F_{\text{exptilt}}$). To apply the EM algorithm as in the single-sample setting, we assume the complete data corresponds to the unobserved failure times corresponding to the right-censoring times from the purely right-censored dataset, as well as the unobserved failure times corresponding to the right-censoring times and left-truncated failure times from the length-biased right-censored dataset. Therefore, the complete-data empirical log-likelihood function is given by:
\[
\ell_c(\alpha, \theta, p) = \sum_{j=1}^{k} \log(p_j) \left\{ r_j + \xi_j + r_j' + \xi_j' + \psi_j' \right\} + (\alpha + \theta h(t_j)) \left\{r_j' + \xi_j' + \psi_j' \right\},
\]
with $r_j = \sum_{i=1}^{n_{\text{RC}}} I(x_i = t_j, \delta_i=1)$, 
$\xi_j = \sum_{i=1}^{n_{\text{RC}}} I(x_i = t_j, \delta_i=0)$, 
$r_j' = \sum_{i=1}^{n_{\text{LBRC}}} I(x_i' = t_j, \delta_i'=1)$, 
$\xi_j' = \sum_{i=1}^{n_{\text{LBRC}}} I(x_i' = t_j, \delta_i'=0)$, 
and $\psi_j' = \sum_{i=1}^{m} I(T_i^* =  t_j)$, where $T_i^*$ corresponds to the unobserved left-truncated failure times. Assuming the right-censored data set is independent of the length-biased right-censored data set, we obtain the conditional expectation of the complete-data log-likelihood given by:
\[
\mathbb{E}(\ell_c(\alpha, \theta, p) | \mathcal{O}) = \sum_{j=1}^{k} \log(p_j) w_{j,1}^{(m)} + \{\alpha + \theta h(t_j)\} w_{j,2}^{(m)},
\]
where
\[
w_{j,1}^{(m)} = r_j + r_j' + \sum_{i=1}^{n_{\text{RC}}} (1 - \delta_i) \frac{\hat{p}_j^{(m)} I(X_i \leq t_j)}{\sum_{j'=1}^{k} \hat{p}_{j'}^{(m)} I(X_i \leq t_{j'})} + \sum_{i=1}^{n_{\text{LBRC}}} (1 - \delta_i') \frac{\hat{q}_j^{(m)} I(X_i' \leq t_j)}{\sum_{j'=1}^{k} \hat{q}_{j'}^{(m)} I(X_i' \leq t_{j'})} + \frac{n_{\text{LBRC}} \hat{q}_j^{(m)}(1 - \frac{t_j}{t_k})}{\hat{\pi}^{(m)}},
\]
and 
\[
w_{j,2}^{(m)} = r_j' + \sum_{i=1}^{n_{\text{LBRC}}} (1 - \delta_i') \frac{\hat{q}_j^{(m)} I(X_i' \leq t_j)}{\sum_{j'=1}^{k} \hat{q}_{j'}^{(m)} I(X_i' \leq t_{j'})} + \frac{n_{\text{LBRC}} \hat{q}_j^{(m)}(1 - \frac{t_j}{t_k})}{\hat{\pi}^{(m)}}.
\]
Thus, the EM algorithm proceeds iteratively by proposing initial estimates $\hat\alpha^{(0)}, \hat\theta^{(0)}$, $\hat p_1^{(0)}, \ldots, \hat p_k^{(0)}$, updating the weights $w_{j,1}$, $w_{j,2}$ using these estimates and then maximizing the conditional expected log-likelihood subject to the constraints $\sum_{j=1}^{k} p_j = 1$ and $\sum_{j=1}^{k} q_j = 1$. 
We remark that compared with the single-sample setting, there is an additional constraint $\sum_{j=1}^{k} q_j = 1$ to ensure the exponentially tilted distribution is proper.
This is equivalent to enforcing $\alpha$'s role as the normalizing constant by imposing the constraint $\alpha = -\log \sum_{j=1}^{k} p_j\exp\{\theta h(t_j)\}$. 

We now show how the M-step is achieved using the Lagrange multiplier method to update $\alpha, \theta$, and $p_j$, $j = 1, \ldots, k$. Let $\blambda=(\lambda_1, \lambda_2)$ be Lagrange multipliers for the two aforementioned constraints, and define the Lagrangian as:
\begin{align*} 
\mathcal{L}(\alpha, \theta, p_1, \ldots, p_k, \blambda) 
& = \mathbb{E}(\ell_c(\alpha, \theta, p) | \mathcal{O}) + \lambda_1 \{ 1-\sum_{j=1}^{k} p_j \} + \lambda_2 \{ 1 - \sum_{j=1}^{k} p_j \exp\{\alpha+\theta h(t_j)\} \} \\ 
& = \left \{ \sum_{j=1}^{k} \log(p_j) w_{j,1}^{(m)} + \{\alpha + \theta h(t_j)\} w_{j,2}^{(m)} \right \} + \lambda_1 \left \{ 1-\sum_{j=1}^{k} p_j \right \} + \lambda_2 \left \{ 1 - \sum_{j=1}^{k} p_j \exp\{\alpha+\theta h(t_j)\} \right \}.
\end{align*} 
The solution of $p_j$ and $(\alpha,\theta)$, which gives the update $(\hat\alpha^{(m+1)}, \hat\theta^{(m+1)}, \hat p_j^{(m+1)})$, to the constrained optimization problem, satisfies 
\[
\frac{\partial \mathcal{L}(\hat\alpha^{(m+1)}, \hat\theta^{(m+1)}, \hat p_1^{(m+1)}, \ldots, \hat p_k^{(m+1)}, \hat \blambda)}{\partial (\alpha, \theta, p_j, \blambda)}=\bsm{0}.
\]
It can be shown that $\hat \lambda_1=\sum_j (w_{j,1}^{(m)}-w_{j,2}^{(m)}) = n_{\text{RC}}$ and $\hat \lambda_2=\sum_j w_{j,2}^{(m)} = n_{\text{LBRC}}/\hat{\pi}^{(m)}$, resulting in the estimator 
\[
\hat p_j^{(m+1)} = \frac{w_{j,1}^{(m)} \hat{\pi}^{(m)}}{n_{\text{RC}} \hat{\pi}^{(m)} + n_{\text{LBRC}} \exp\{\hat\alpha^{(m+1)}+\hat\theta^{(m+1)} h(t_j)\}}.
\]
where, to avoid notational confusion, we note that $\hat p_j^{(m+1)} = p_j^{(m+1)}(\hat\alpha^{(m+1)},\hat\theta^{(m+1)})$.
The updated values of $\hat\alpha^{(m+1)}$ and $\hat\theta^{(m+1)}$ are obtained by setting the partial derivatives of the Lagrangian, with respect to $(\alpha,\theta)$, to zero and then solving the corresponding equations. 
Equivalently, they can be found by maximizing the expected complete-data log-likelihood $\mathbb{E}(\ell_c(\alpha, \theta, p) | \mathcal{O})$, after substituting
\[
p_j=p_j^{(m+1)}(\alpha,\theta) \coloneqq 
\frac{w_{j,1}^{(m)} \hat{\pi}^{(m)}}{n_{\text{RC}} \hat{\pi}^{(m)} + n_{\text{LBRC}} \exp\{\alpha+\theta h(t_j)\}},
\]
yielding the objective function:
\[
\sum_{j=1}^{k} \log\{p_j^{(m+1)}(\alpha,\theta)\} w_{j,1}^{(m)} + \{\alpha + \theta h(t_j)\} w_{j,2}^{(m)},
\]
to be maximized with respect to $(\alpha,\theta)$.
\com{
This objective function is the profile empirical log-likelihood for $(\alpha,\theta)$ at the $(m+1)$th M-step. It is obtained by profiling out the probability masses $p_1,\ldots,p_k$ under the normalization constraints $\sum_{j=1}^k p_j=1$ and $\sum_{j=1}^k p_j\exp\{\alpha+\theta h(t_j)\}=1$. Substituting the resulting profiled masses $p_j^{(m+1)}(\alpha,\theta)$ back into the expected complete-data empirical log-likelihood yields the displayed profile objective function, which is then maximized with respect to $(\alpha,\theta)$. Thus, although the empirical likelihood is embedded within an EM algorithm because of censoring and length bias, the M-step retains the profile empirical likelihood structure.
}

\com{After profiling out the probability masses, the resulting optimization over $(\alpha,\theta)$ is unconstrained and is thus easily implementable.}
\com{Using the arguments of \cite{vardi}, the EM algorithm ascends at each step and converges to a maximizer of the likelihood function. We note, however, that unlike the single-sample setting of \cite{vardi}, \cite{david} show that in certain settings where multiple biased samples are combined, the maximizer is not necessarily unique. Interestingly, in the case where right-censored data are drawn from the reference distribution and the underlying failure times in the length-biased right-censored data are drawn from the exponentially tilted distribution, the maximizer exists, is unique and consistent as $n_{\text{RC}} + n_{\text{LBRC}} \rightarrow \infty$. We formalize these properties in the theorem below.}
\begin{theorem}
\com{Let $\hat{\alpha}$, $\hat{\theta}$ and $\hat{F}_0$ denote the maximizer of $L(\alpha, \theta, F_0)$ given by \eqref{2samplike}. Suppose the right-censored sample failure/censoring times are independent of the length-biased right-censored failure/censoring times and that $\hat{\beta} = n_{\text{RC}}/(n_{\text{RC}} + n_{\text{LBRC}})
\rightarrow \beta \in (0, 1)$ as $n_{\text{RC}}, n_{\text{LBRC}} \rightarrow \infty$. Then, the maximizers $\hat{\alpha}$, $\hat{\theta}$ and $\hat{F}_0$ exist, are unique and $\hat{\theta} \rightarrow \theta_0$, $\hat{\alpha} \rightarrow \alpha_0$ and $\hat{F}_0 \rightarrow F_0$ uniformly for all $t \in (0, \tau_0)$ as $n_{\text{RC}} + n_{\text{LBRC}} \rightarrow \infty$.}
\label{Theorem1}
\end{theorem}
\com{When $\alpha = \theta = 0$, \cite{mcvit2} established the asymptotic properties of the NPMLE of $\mathrm{d}F_0$ for a combined set of right-censored and length-biased right-censored failure time data (i.e. the underlying failure times of both sets of data are drawn from the same reference distribution). The proof of Theorem \ref{Theorem1} follows a similar approach to \cite{mcvit2} based on the proofs of \cite{zhu2} for a single sample of left-truncated right-censored data drawn from an exponentially tilted distribution. For the proof of Theorem \ref{Theorem1}, see the Supplementary Materials.}

\subsection{$K > 2$ Sample Methods}

When more than two samples are combined, the procedure discussed above in the two-sample setting can easily be applied. 
First, the distribution of one of the (unbiased) samples is chosen to be the reference distribution $F_0$.
Then, the distributions of the other $K-1$ samples are modelled under a DRM that shares the same reference distribution $F_0$ and basis function $h(\cdot)$, but may have different exponential tilting parameters $\theta_1, \ldots, \theta_{K-1}$.
Finally, within the EM framework, one can form the complete-data empirical log-likelihood, compute the expectations at each time point in the E-step, and maximize the log-likelihood in the M-step.
Iterating these two steps until convergence yields estimators of the unknown parameters and the survival functions for all the samples.

\section{Simulation Study}
\label{sec:simulation}

To assess the performance of the proposed DRM estimation procedure for combined right-censored (RC) and length-biased right-censored (LBRC) failure time data, we generated simulated failure time data. Specifically, we assumed the right-censored failure times were distributed according to a gamma distribution with a varying shape parameter equal to $0.5$, $1$ and $2$ and a fixed scale parameter equal to $2$. For each of the shape parameter cases, the underlying failure times in the LBRC data were generated according to a gamma distribution with shape parameter equal to $1.5$, $2$ and $3$, respectively, and scale parameter equal to $2$. Thus, under the DRM framework, \com{the underlying failure times in the RC data were drawn from the reference distribution and} the \com{underlying} failure times in the LBRC data were drawn from a DRM with $\theta = 1$ and $h(x) = \log(x)$ \com{(i.e. the exponentially tilted distribution)}. The RC failure times were right-censored by an independent exponentially distributed random variable to allow for $15\%$ and $30\%$ right-censoring. \com{We note that the assumption of the right-censoring times being exponentially distributed is not necessary in the empirical analyses of the proposed methodology; however, its simplicity ensures control in tuning the proportion of censoring in both the unbiased and biased samples. For a discussion on the selection of censoring distribution parameters to obtain exact censoring proportions for simple failure time distributions, see \cite{wan}. Furthermore, we assumed that the underlying failure times were gamma distributed to easily sample from a reference gamma distribution and an exponentially tilted gamma distribution since the gamma distribution is closed under polynomial weights. That is, it is straightforward to show that if $f(x; \theta)$ is a gamma density function, then for positive integers $k$,  $x^kf(x;\theta)/\left[\int_{0}^{\infty}u^kf(u;\theta)\mathrm{d}u\right]$ will also be a gamma density function. These distributional assumptions on the censoring times and failure times are not necessary for the proposed methodologies and were only used to facilitate straightforward sampling of the right-censored data drawn from the failure time reference density function and the length-biased right-censored data drawn from the exponentially tilted failure time density function.} The left-truncation times for the LBRC data were generated from a uniform distribution over the interval $(0, 50)$ where the forward failure times were right-censored by an exponentially distributed random variable to allow for $15\%$ and $30\%$ right-censoring. \com{Thus, to generate an observation in the LBRC data subset, we sampled a failure time from the exponentially tilted failure time distribution and only retained it if it surpassed the sampled left-truncation time. If so, we then right-censored the residual failure time (i.e. the difference of the sampled failure time less the sampled left-truncation time) by an independent exponentially distributed random variable and retained both the minimum and censoring indicator.} For the different data sets, we generated samples consisting of all combinations in sizes of $50, 100$ and $200$ observations for the RC and LBRC data. \com{In the two subsections that follow, we compare the performance of the individual data set survival function estimators to the combined data DRM estimator and combined data survival function NPMLE. In Subsection \ref{sim1subsec}, we consider three different reference failure time distributions with increasing, decreasing or constant hazard rates where the DRM is correctly specified. In Subsection \ref{sim2subsec}, we allow for possible overspecification or misspecification of the DRM under the same failure time distribution settings.}

\subsection{\com{Individual and Combined Data Estimator Comparisons}}
\label{sim1subsec}

Using the simulated data sets, we compared the performance of various estimators based on their empirical Kolmogorov--Smirnov (KS) distances: 
$$KS(\hat{F}, F) = \max_{x: x\in\{x_1, \ldots, x_k\}} \big|\hat{F}(x) - F(x)\big|$$
where the points $x_1, \ldots, x_k$ were selected to cover the majority of the support of the different gamma distributions. We considered the following estimators: (i) Kaplan--Meier (KM) product-limit estimator (RC data only) \citep{kapla}, \com{(ii)} Survival Function NPMLE (LBRC data only) \citep{asgha2}, \com{(iii)} Survival Function NPMLE (Combined RC and LBRC data assuming a common distribution) \citep{mcvit, mcvit2}, and \com{(iv)} Our proposed DRM estimator (Combined RC and LBRC data). The Kaplan--Meier estimator in (i) is based only on the RC data. The survival function NPMLE in \com{(ii)} uses the uniformity of the left-truncation times in its estimation procedures for only the LBRC data, whereas the estimator in \com{(iii)} is an extension of the NPMLE in \com{(ii)} that allows for the inclusion of RC data with the LBRC data. Tables~\ref{corr_spec_dechaz}, \ref{corr_spec_consthaz} and \ref{corr_spec_inchaz} report the mean empirical KS distances and their standard deviations over 1,000 simulation iterations for the \com{four} estimators \com{(see Tables~\ref{corr_spec_consthaz} and~\ref{corr_spec_inchaz} in Appendix)}.

Overall, the DRM estimator tends to yield the smallest mean KS distance, and generally the smallest standard deviation, across nearly all scenarios, whereas \com{the} NPMLE (RC+LBRC) \com{tends} to have the largest KS distances. 
\com{As the NPMLE (RC+LBRC) estimator assumes a common underlying failure time distribution for the RC and LBRC samples, it targets a single common survival function when this assumption is false in our simulation setting. By contrast, the DRM estimator allows the underlying failure time distribution of the LBRC data to be an exponentially tilted version of the RC failure time reference distribution. Therefore, the DRM can borrow information across the two samples while still allowing distributional differences between them, which explains its improved performance in the heterogeneous settings considered here.}

\subsection{\com{Overspecified and Misspecified Estimator Comparisons}}
\label{sim2subsec}

We conducted additional simulations to assess the robustness of the DRM under \com{three different scenarios. In Scenario (i), we assumed an overspecified DRM} with a vector basis $h(x)=(x, \log(x))$ when the correct basis function was only $h^*(x) = \log(x)$, and \com{in Scenario (ii), we assumed a misspecified DRM} with $h(x)=\sqrt{x}$ or $h(x)=x$. 
In these settings, the DRM generally continues to produce the smallest KS distances; see Tables S1--S9 in the Supplementary Materials for specific results in these cases.
We note that the overspecified model in \com{Scenario} (i) covers the gamma family without the equal-scale-parameter assumption.
\com{In Scenario (iii), we examined} the case where the RC and LBRC \com{underlying} failure-time distributions are truly identical (gamma with shape parameter $0.5$, $1$, or $2$ and scale parameter $2$).  
Fitting the DRM with $h(x) = \log(x)$ to samples that are ten times larger than above, our proposed method performs essentially as well as the pooled RC+LBRC NPMLE, indicating that with sufficient information, the DRM is able to adaptively determine from the data the equivalence or difference between the unbiased and exponentially tilted distributions. For detailed Kolmogorov--Smirnov results, refer to Tables S10--S12 in the Supplementary Materials. 
\com{Thus, in applied settings where the two data distributions differ, the DRM offers a clear advantage over existing methods. When the two underlying failure time distributions are indeed identical, this setting is nested within the DRM framework as the special case of $(\alpha, \theta)=\boldsymbol{0}$, and the results in Scenario (iii) show that the DRM performs comparably to the combined data NPMLE, which suggests little loss of efficiency in this common distribution setting.}

\begin{table}[!htbp] \centering 
  \caption{\com{Average empirical KS distances (standard deviations in brackets) for Kaplan--Meier estimator using only RC data (KM (RC)), the survival function NPMLE using only LBRC data (NPMLE (LBRC)), the survival function NPMLE using both RC and LBRC data (NPMLE RC+LBRC) and the density ratio model estimator (DRM). Failure times are gamma distributed with decreasing hazard with varying censoring proportions and sample sizes for RC and LBRC data (RC \%/SS, LBRC \%/SS, respectively).}} 
  \label{corr_spec_dechaz} 
  {   \renewcommand{\arraystretch}{1.5}   \resizebox{\textwidth}{!}{
\begin{tabular}{cccc|cccc} 
RC \% & LBRC \% & RC SS & LBRC SS & KM (RC) & NPMLE (LBRC) & NPMLE (RC+LBRC) & DRM \\ \hline
$15$ & $15$ & $50$ & $50$ & $0.123$ (0.0355) &  $0.193$ (0.0747) & $0.141$ (0.0392) & $0.115$ (0.0368) \\ 
$30$ & $15$ & $50$ & $50$ & $0.145$ (0.0412) &  $0.192$ (0.0742) & $0.151$ (0.0424) & $0.124$ (0.0398) \\ 
$15$ & $15$ & $100$ & $50$ & $0.089$ (0.0270) & $0.192$ (0.0713) & $0.103$ (0.0279) & $0.085$ (0.0276) \\ 
$30$ & $15$ & $100$ & $50$ & $0.102$ (0.0286) & $0.191$ (0.0782) & $0.115$ (0.0318) & $0.090$ (0.0290) \\ 
$15$ & $15$ & $200$ & $50$ & $0.062$ (0.0179) & $0.192$ (0.0717) & $0.074$ (0.0197) & $0.060$ (0.0182) \\ 
$30$ & $15$ & $200$ & $50$ & $0.074$ (0.0197) & $0.193$ (0.0695) & $0.085$ (0.0219) & $0.066$ (0.0194) \\ \hline \hline
$15$ & $30$ & $50$ & $50$ & $0.125$ (0.0368) &  $0.199$ (0.0746) & $0.142$ (0.0402) & $0.117$ (0.0372) \\ 
$30$ & $30$ & $50$ & $50$ & $0.145$ (0.0414) &  $0.198$ (0.0736) & $0.152$ (0.0444) & $0.124$ (0.0396) \\ 
$15$ & $30$ & $100$ & $50$ & $0.088$ (0.0270) & $0.199$ (0.0770) & $0.106$ (0.0291) & $0.084$ (0.0274) \\ 
$30$ & $30$ & $100$ & $50$ & $0.101$ (0.0278) & $0.197$ (0.0699) & $0.113$ (0.0294) & $0.088$ (0.0269) \\ 
$15$ & $30$ & $200$ & $50$ & $0.062$ (0.0186) & $0.197$ (0.0798) & $0.074$ (0.0191) & $0.060$ (0.0186) \\ 
$30$ & $30$ & $200$ & $50$ & $0.073$ (0.0208) & $0.195$ (0.0717) & $0.085$ (0.0226) & $0.066$ (0.0199) \\ \hline \hline
$15$ & $15$ & $50$ & $100$ & $0.122$ (0.0372) & $0.143$ (0.0522) & $0.165$ (0.0382) & $0.113$ (0.0371) \\ 
$30$ & $15$ & $50$ & $100$ & $0.143$ (0.0401) & $0.141$ (0.0527) & $0.173$ (0.0408) & $0.118$ (0.0391) \\ 
$15$ & $15$ & $100$ & $100$ & $0.087$ (0.0261) & $0.142$ (0.0514) & $0.124$ (0.0301) & $0.082$ (0.0272) \\ 
$30$ & $15$ & $100$ & $100$ & $0.103$ (0.0289) & $0.141$ (0.0547) & $0.135$ (0.0304) & $0.087$ (0.0278) \\ 
$15$ & $15$ & $200$ & $100$ & $0.062$ (0.0184) & $0.141$ (0.0582) & $0.089$ (0.0197) & $0.059$ (0.0188) \\ 
$30$ & $15$ & $200$ & $100$ & $0.073$ (0.0205) & $0.142$ (0.0509) & $0.102$ (0.0223) & $0.063$ (0.0202) \\ \hline \hline 
$15$ & $30$ & $50$ & $100$ & $0.125$ (0.0377) & $0.146$ (0.0570) & $0.164$ (0.0398) & $0.115$ (0.0399) \\ 
$30$ & $30$ & $50$ & $100$ & $0.144$ (0.0423) & $0.146$ (0.0591) & $0.172$ (0.0436) & $0.121$ (0.0387) \\ 
$15$ & $30$ & $100$ & $100$ & $0.088$ (0.0257) & $0.146$ (0.0564) & $0.124$ (0.0290) & $0.083$ (0.0269) \\ 
$30$ & $30$ & $100$ & $100$ & $0.101$ (0.0272) & $0.146$ (0.0545) & $0.135$ (0.0300) & $0.087$ (0.0273) \\ 
$15$ & $30$ & $200$ & $100$ & $0.062$ (0.0188) & $0.144$ (0.0518) & $0.089$ (0.0202) & $0.059$ (0.0194) \\ 
$30$ & $30$ & $200$ & $100$ & $0.073$ (0.0197) & $0.146$ (0.0551) & $0.101$ (0.0222) & $0.062$ (0.0188) \\ \hline \hline
$15$ & $15$ & $50$ & $200$ & $0.125$ (0.0390) & $0.105$ (0.0467) & $0.194$ (0.0378) & $0.111$ (0.0392) \\ 
$30$ & $15$ & $50$ & $200$ & $0.144$ (0.0427) & $0.105$ (0.0436) & $0.203$ (0.0410) & $0.114$ (0.0040) \\ 
$15$ & $15$ & $100$ & $200$ & $0.086$ (0.0253) & $0.105$ (0.0414) & $0.152$ (0.0272) & $0.078$ (0.0256) \\ 
$30$ & $15$ & $100$ & $200$ & $0.102$ (0.0294) & $0.107$ (0.0449) & $0.163$ (0.0300) & $0.084$ (0.0301) \\ 
$15$ & $15$ & $200$ & $200$ & $0.062$ (0.0186) & $0.104$ (0.0397) & $0.114$ (0.0200) & $0.058$ (0.0189) \\ 
$30$ & $15$ & $200$ & $200$ & $0.073$ (0.0201) & $0.104$ (0.0407) & $0.125$ (0.0215) & $0.060$ (0.0201) \\ \hline \hline 
$15$ & $30$ & $50$ & $200$ & $0.123$ (0.0370) & $0.106$ (0.0386) & $0.194$ (0.0400) & $0.109$ (0.0382) \\ 
$30$ & $30$ & $50$ & $200$ & $0.142$ (0.0395) & $0.107$ (0.0428) & $0.204$ (0.0397) & $0.112$ (0.0385) \\ 
$15$ & $30$ & $100$ & $200$ & $0.089$ (0.0271) & $0.107$ (0.0407) & $0.151$ (0.0285)  & $0.081$ (0.0274) \\ 
$30$ & $30$ & $100$ & $200$ & $0.103$ (0.0293) & $0.104$ (0.0392) & $0.162$ (0.0303)  & $0.085$ (0.0287) \\ 
$15$ & $30$ & $200$ & $200$ & $0.062$ (0.0190) & $0.105$ (0.0410) & $0.113$ (0.0195)  & $0.058$ (0.0190) \\ 
$30$ & $30$ & $200$ & $200$ & $0.072$ (0.0201) & $0.105$ (0.0381) & $0.125$ (0.0217) & $0.061$ (0.0200) \\ \hline
\end{tabular} } }
\end{table}

\section{Application}
\label{sec:application}

The analysis of the length of time spent in hospitals from admission to discharge is key in adjusting hospital policy, resource allocation and improving the level of care given to the patients. The Population Health Records platform is a web application which combines health data from multiple sources related to population health monitoring \citep{shaba}. Using this platform, we considered a 30-day (one-month) period for admission/discharge dates for patients in a Montreal-area hospital. Due to various privacy constraints on the type of data, the admission/discharge dates were listed on an integer scale with the earliest date set to $0$ (i.e. specific calendar date information removed). Individuals who were admitted before the one-month period but were still in the hospital during this period formed the length-biased right-censored data subset, whereas those individuals who were admitted during the one-month period formed the right-censored data subset. For both sets of data, the durations were right-censored by the end date of the one-month period. The RC data consists of 369 observations with a right-censoring proportion of approximately 20\%, whereas the LBRC data consists of 69 observations with a right-censoring proportion of approximately 28\%. \com{We assessed the stationarity assumption in the set of left-truncated right-censored failure time data graphically using the methods described by \cite{asgha3} and found that the stationarity assumption was plausible. For the specific graphs and associated details, see the Supplementary Materials.}

\begin{figure}
\centering
    \includegraphics[scale=0.5]{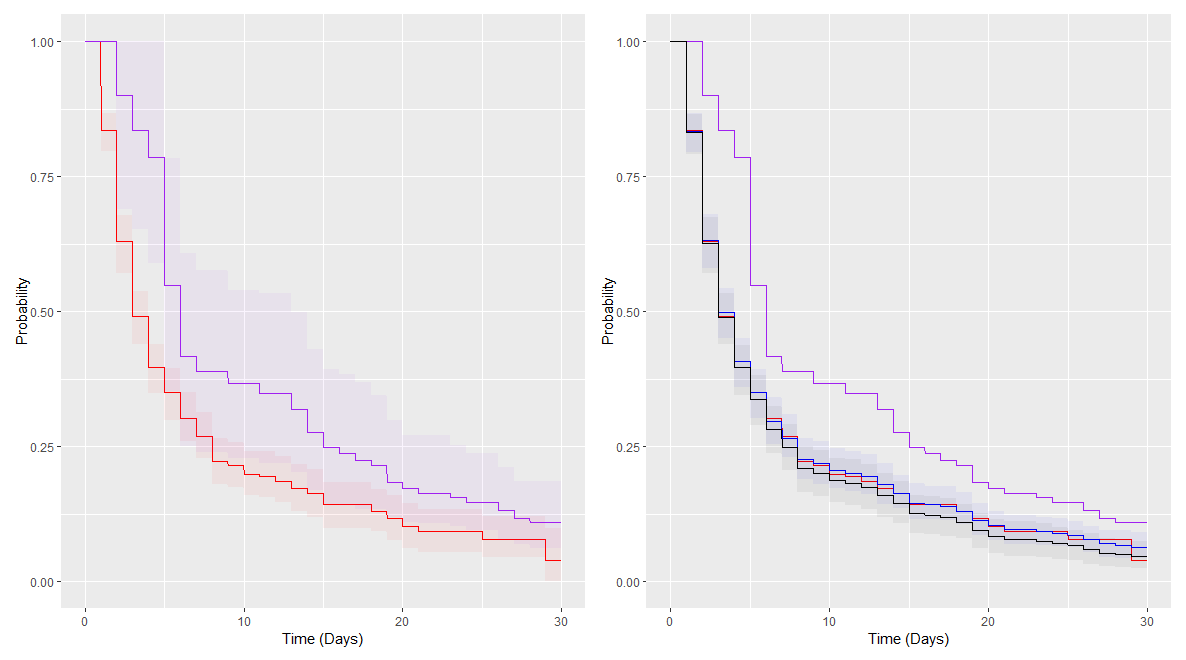}
    \caption{Estimated survival curves (solid lines) and 95\% pointwise confidence intervals (shaded areas) for the duration of time spent in a Montreal-area hospital using only the right-censored durations (KM estimate (red)), only the length-biased right-censored durations (NPMLE (purple)) \com{and} both the right-censored and length-biased right-censored durations (NPMLE (blue), DRM (black)). \com{Left panel - individual data set estimates and confidence intervals, Right panel - individual data set estimates and combined data set estimates and confidence intervals}.}
    \label{Fig2}
\end{figure}

First, using only the subset of RC data, we computed the Kaplan--Meier estimator of the survival function and using only the subset of LBRC data, we computed the NPMLE of the survival function. \com{Using a nonparametric bootstrapping procedure, we computed 95\% pointwise confidence intervals for the estimates and present the results in the left panel of Figure \ref{Fig2}. From the estimated survival curves, there are clear differences in the shapes of the hospital stay duration distributions for the RC and LBRC data.} Based on this observation, we believe the use of the DRM is sensible to capture the similarities in the duration distributions, where we assume the RC observations are drawn from the reference distribution. 

Using the \com{single-sample KM estimator and NPMLE for the RC and LBRC data, respectively, as well as the NPMLE for combined RC and LBRC data}, we computed the survival function estimates and recorded the 25\%, 50\% and 75\% quantiles. For the DRM model, \com{we considered $h(x)\in \{\log(x),\sqrt{x},x,x^2\}$ as a set of commonly used elementary transformations to examine different possible exponential tilts.
The $\log(x)$ choice is motivated by the classical length-biased form, while $\sqrt{x}$, $x$, and $x^2$ allow different directions of departure from the reference distribution.}
However, in all cases, the estimated quantiles were essentially equal, \com{suggesting that the main conclusions of this application are not sensitive to these choices of $h(x)$.
Across these choices, the largest estimated value of $\theta$ was $0.3245$}. 
In Table~\ref{real_quan}, we report the quantile estimates for the various non-DRM estimators and for the DRM estimator with $h(x) = \log(x)$. Using only the right-censored failure time data, the estimated median was approximately 3 days, whereas using \com{the survival function NPMLE for} only the LBRC data, the \com{estimated} median was approximately 6 days. When the RC and LBRC data were combined, the NPMLE using both the RC and LBRC data tended to agree with the DRM estimates with the function $h(x) = \log(x)$ in producing a median estimate of $3$ days. We note that given the relative sizes of the RC and LBRC data sets, the DRM and NPMLE (RC+LBRC) estimates are heavily influenced by the values of the RC data and thus generally agree with the estimates provided by the Kaplan--Meier estimator. Additionally, due to the privacy constraints on the analyzed data, we were unable to informatively select a function $h(x)$ based on the observed data. In practice, if the introduction of a particular hospital policy or procedure is known to lengthen the duration of stay within the hospital, the statistical analyst may select higher-order polynomial terms for $h(x)$ to capture this behaviour.

Finally, using the nonparametric bootstrapping procedure of \cite{efron} and resampling with replacement from both the RC and LBRC data sets, we computed 95\% \com{pointwise} confidence intervals for the \com{survival function NPMLE and DRM estimates} using 150 bootstrapped samples. \com{We plot the individual data set survival function estimates from the left-panel as well as the combined data NPMLE and DRM estimates with pointwise confidence intervals in the right-panel of Figure~\ref{Fig2}.} \com{While it is possible to derive uniform confidence bands for the survival function using right-censored and left-truncated right-censored failure time data separately, the extension to combined cohort data remains an open problem. For recent references see \citep{sachs, shari}.} The 95\% confidence interval for $\theta$ was given by $(0.040, 0.720)$, suggesting no evidence at the $0.05$ level that the RC data failure time distribution was the same as the unbiased LBRC data failure time distribution. Additionally, from the plot of the bootstrapped confidence intervals of the survival function, the DRM estimation procedure tends to yield relatively narrow intervals across all time points and exhibits similar confidence interval size to the NPMLE for both RC and LBRC data. However, the difference observed between the DRM and NPMLE survival curves is mainly attributable to the evidence that $\theta \neq 0$, indicating that the RC and LBRC samples do not follow identical underlying distributions.

\begin{table} \centering 
  \caption{Comparison of 25\%, 50\% and 75\% quantiles for the length of stay (measured in days) in a Montreal-area hospital for data collected over a one month period based on the Kaplan--Meier estimator using only the RC data (KM (RC)), the survival function NPMLE using only the LBRC data (NPMLE (LBRC)), the survival function NPMLE using both RC and LBRC data (NPMLE RC+LBRC) and the density ratio model estimator (DRM) with $h(x) = \log(x)$.} 
  \label{real_quan}
\begin{tabular}{c|ccc}
 Estimator & 25\% Quantile & 50\% Quantile & 75\% Quantile \\ \hline
 KM (RC) & 2 & 3 & 8 \\
 NPMLE (LBRC) & 5 & 6 & 15 \\
 NPMLE (RC+LBRC) & 2 & 3 & 8 \\ 
 DRM (RC+LBRC) & 2 & 3 & 7 \\
\end{tabular}
\end{table}

\section{Discussion}
\label{sec:discussion}

In this paper, we presented a semiparametric modelling procedure, via the density ratio model, to jointly analyze a combined data set comprised of both randomly right-censored failure times and length-biased right-censored failure times. 
In particular, we showed how the EM algorithm for the NPMLE of the survival function using only length-biased right-censored data could be adapted to this combined data setting, where the right-censored failure times were assumed to be drawn from the reference density function and the underlying failure times in the length-biased right-censored data set were assumed to be drawn from an exponentially tilted version of the reference density. 
We assessed the performance of the proposed methodology on simulated failure time data and found that the DRM outperforms the other methods in the majority of cases. 
We then applied the DRM to model the length of stay in a Montreal-area hospital, measured from admission to discharge. 

In our EM algorithm, the E-step computes the expected number of failures at each observed failure/censoring time, and the M-step updates the empirical likelihood estimators of the model parameters and the probability masses of the reference distribution accordingly. 
Our EM procedure imposes no assumptions on the form of the censoring distribution or on the realized values of the censoring times. By contrast, the inference approach of \cite{cai}, developed for multiple censored samples, treats censoring times as known fixed constants. That restrictive assumption enables estimation of the unknown censoring probability from the empirical distribution of the observed censoring times and leads to a partial empirical likelihood. In this paper, we allow both fixed and random censoring times throughout, and our EL-EM framework remains valid without specifying or fixing the censoring distribution.

We also note an interesting connection between exponentially-tilted density functions and the conditional failure time density function for left-truncated data. Specifically, let $f_T$ denote the marginal density of the failure time $T$ and $G_A$ denote the cumulative distribution function of the left-truncation random variable $A$. Then, the conditional density of the observed failure time $T$ is given by:
\[
f(t|T>A) = \frac{f_T(t)G_A(t)}{\int f_T(u)G_A(u) \mathrm{d}u}.
\]
If $G_A(t)$ is specified by a parametric model $G_A(t; \phi)$ and is log-linear in its parameters (or an appropriate transformation), i.e., $\log \{G_A(t; \phi)\} = \alpha + \theta h(t)$ for unknown $\theta$, then the conditional density given above simplifies to a DRM:
\begin{align*}
f(t|T>A) & = f_T(t) \exp\left( -\log\left\{\int f_T(u)G_A(u; \phi) \mathrm{d}u\right\} + \log\{G_A(t; \phi)\} \right) \\ 
& = f_T(t) \exp\{ \alpha^* + \theta h(t) \},
\end{align*}
where $\alpha^*=\alpha-\log\{\int f_T(u)G_A(u; \phi) \mathrm{d}u\}$ does not depend on $t$ and can be treated as an unknown normalizing parameter.
Thus, any left-truncated density with a parametrically specified log-linear truncation distribution can be expressed as an exponentially tilted density. 
However, the converse is not necessarily true, as not every choice of the basis function $h(x)$ in the exponential tilt yields the logarithm of a cumulative distribution function. 
A systematic treatment of this observation is left for future work on left-truncated failure time data.
While the preceding discussion implicitly assumes that $G_A(t)>0$, which fails when the truncation variable $A$ has a bounded support, for example, when $A$ is uniformly distributed over a finite interval, the same argument may still apply on the common support of $A$ and $T$.

Another area for future research is the inclusion of covariate data in the DRM for different types of partially observed failure time data. For the standard case of i.i.d. data, the inclusion of covariates severely complicates the resulting estimation procedures, with few extensions to the cases with partially observed or biased failure time data. For additional discussion on the inclusion of covariate data, see \citep{luo, diao2012maximum, huang2014joint, zhu, zhang2}. 

Finally, throughout all the proposed estimation procedures, we assumed that the right-censored failure times were drawn directly from the reference distribution and the length-biased right-censored failure times were drawn from an exponentially tilted version of the reference density via the DRM. 
This assumption is equivalent to assuming both samples are drawn from exponentially tilted versions of a common reference distribution $F_{\text{ref}}$ with a common basis function $h(x)$:
\begin{align}
\mathrm{d}F_{\text{RC}}(x; \theta) = \exp\{\alpha_{\text{RC}} + \theta_{\text{RC}} h(x)\} \mathrm{d}F_{\text{ref}}(x),
\hspace{5mm}
\mathrm{d}F_{\text{LBRC}}(x; \theta) = \exp\{\alpha_{\text{LBRC}} + \theta_{\text{LBRC}} h(x)\} \mathrm{d}F_{\text{ref}}(x).
\label{DRM_equiv}
\end{align}
which corresponds to case (iii) in Section 2.2.
However, in this case, as no data are observed directly from the reference distribution $F_{\text{ref}}$, additional moment constraints on $h(X)$ with respect to $F_{\text{ref}}$ are needed to identify the model components in \eqref{DRM_equiv} (see \citet{zhang2}).
Moreover, inference under this formulation essentially collapses to case (ii) considered in this paper, where one of the observed distributions is taken as the reference, while incurring unnecessary computation. 
Therefore, in the absence of covariates, it is more straightforward to select one observed distribution as the reference and treat the others as its exponentially tilted versions; the DRM is invariant to which observed distribution is chosen as the reference.

\newpage 

\section*{\com{Appendix}}
\begin{table}[H] \centering 
  \caption{\com{Average empirical KS distances (standard deviations in brackets) for Kaplan--Meier estimator using only RC data (KM (RC)), the survival function NPMLE using only LBRC data (NPMLE (LBRC)), the survival function NPMLE using both RC and LBRC data (NPMLE RC+LBRC) and the density ratio model estimator (DRM). Failure times are gamma distributed with constant hazard with varying censoring proportions and sample sizes for RC and LBRC data (RC \%/SS, LBRC \%/SS, respectively).}} 
  \label{corr_spec_consthaz} 
  {   \renewcommand{\arraystretch}{1.2}   \resizebox{\textwidth}{!}{
\begin{tabular}{cccc|cccc} 
RC \% & LBRC \% & RC SS & LBRC SS & KM (RC) & NPMLE (LBRC) & NPMLE (RC+LBRC) & DRM \\ \hline
$15$ & $15$ & $50$ & $50$ & $0.125$ (0.0385) & $0.183$ (0.0666) & $0.150$ (0.0427) & $0.116$ (0.0389) \\ 
$30$ & $15$ & $50$ & $50$ & $0.139$ (0.0407) & $0.184$ (0.0713) & $0.158$ (0.0441) & $0.124$ (0.0400) \\ 
$15$ & $15$ & $100$ & $50$ & $0.089$ (0.0268) & $0.182$ (0.0723) & $0.108$ (0.0315) & $0.084$ (0.0271) \\ 
$30$ & $15$ & $100$ & $50$ & $0.098$ (0.0283) & $0.180$ (0.0645) & $0.116$ (0.0333) & $0.089$ (0.0280) \\ 
$15$ & $15$ & $200$ & $50$ & $0.063$ (0.0195) & $0.180$ (0.0669) & $0.075$ (0.0214) & $0.061$ (0.0196) \\ 
$30$ & $15$ & $200$ & $50$ & $0.070$ (0.0204) & $0.180$ (0.0652) & $0.083$ (0.0233) & $0.066$ (0.0203) \\ \hline \hline
$15$ & $30$ & $50$ & $50$ & $0.127$ (0.0379) & $0.171$ (0.0603) & $0.149$ (0.0426) & $0.116$ (0.0372) \\ 
$30$ & $30$ & $50$ & $50$ & $0.138$ (0.0398) & $0.172$ (0.0599)  & $0.157$ (0.0435) & $0.121$ (0.0401) \\ 
$15$ & $30$ & $100$ & $50$ & $0.089$ (0.0257) & $0.176$ (0.0671) & $0.107$ (0.0301) & $0.084$ (0.0260) \\ 
$30$ & $30$ & $100$ & $50$ & $0.099$ (0.0292) & $0.174$ (0.0638) & $0.116$ (0.0331) & $0.090$ (0.0292) \\ 
$15$ & $30$ & $200$ & $50$ & $0.062$ (0.0189) & $0.176$ (0.0672) & $0.074$ (0.0207) & $0.060$ (0.0191) \\ 
$30$ & $30$ & $200$ & $50$ & $0.069$ (0.0206) & $0.173$ (0.0642) & $0.085$ (0.0234) & $0.064$ (0.0209) \\ \hline \hline 
$15$ & $15$ & $50$ & $100$ & $0.124$ (0.0372) & $0.132$ (0.0512) & $0.172$ (0.00407) & $0.110$ (0.0376) \\ 
$30$ & $15$ & $50$ & $100$ & $0.137$ (0.0404) & $0.131$ (0.0447) & $0.181$ (0.0423) & $0.116$ (0.0388) \\ 
$15$ & $15$ & $100$ & $100$ & $0.089$ (0.0274) & $0.130$ (0.0502) & $0.129$ (0.0305) & $0.082$ (0.0269) \\ 
$30$ & $15$ & $100$ & $100$ & $0.098$ (0.0275) & $0.129$ (0.0470) & $0.138$ (0.0326) & $0.086$ (0.0281) \\ 
$15$ & $15$ & $200$ & $100$ & $0.062$ (0.0182) & $0.128$ (0.0453) & $0.093$ (0.0230) & $0.059$ (0.0184) \\ 
$30$ & $15$ & $200$ & $100$ & $0.068$ (0.0191) & $0.132$ (0.0478) & $0.101$ (0.0229) & $0.062$ (0.0190) \\ \hline \hline
$15$ & $30$ & $50$ & $100$ & $0.124$ (0.0365) & $0.128$ (0.0469) & $0.171$ (0.0398) & $0.109$ (0.0358) \\ 
$30$ & $30$ & $50$ & $100$ & $0.138$ (0.0400) & $0.127$ (0.0476) & $0.182$ (0.0415) & $0.114$ (0.0390) \\ 
$15$ & $30$ & $100$ & $100$ & $0.089$ (0.0274) & $0.125$ (0.0466) & $0.131$ (0.0298) & $0.081$ (0.0267) \\ 
$30$ & $30$ & $100$ & $100$ & $0.098$ (0.0286) & $0.127$ (0.0458) & $0.140$ (0.0325) & $0.085$ (0.0284) \\ 
$15$ & $30$ & $200$ & $100$ & $0.064$ (0.0188) & $0.129$ (0.0469) & $0.095$ (0.0230) & $0.060$ (0.0194) \\ 
$30$ & $30$ & $200$ & $100$ & $0.069$ (0.0199) & $0.126$ (0.0468) & $0.102$ (0.0233) & $0.063$ (0.0204) \\ \hline \hline 
$15$ & $15$ & $50$ & $200$ & $0.125$ (0.0378) & $0.096$ (0.0319) & $0.208$ (0.0381) & $0.104$ (0.0362) \\ 
$30$ & $15$ & $50$ & $200$ & $0.139$ (0.0421) & $0.094$ (0.0327) & $0.212$ (0.0394) & $0.110$ (0.0386) \\ 
$15$ & $15$ & $100$ & $200$ & $0.089$ (0.0277) & $0.094$ (0.0335) & $0.160$ (0.0288) & $0.078$ (0.0263) \\ 
$30$ & $15$ & $100$ & $200$ & $0.098$ (0.0283) & $0.095$ (0.0347) & $0.169$ (0.0304) & $0.082$ (0.0276) \\ 
$15$ & $15$ & $200$ & $200$ & $0.063$ (0.0191) & $0.093$ (0.0307) & $0.120$ (0.0220) & $0.058$ (0.0193) \\ 
$30$ & $15$ & $200$ & $200$ & $0.070$ (0.0199) & $0.094$ (0.0333) & $0.128$ (0.0223) & $0.061$ (0.0193) \\ \hline \hline
$15$ & $30$ & $50$ & $200$ & $0.127$ (0.0369) & $0.091$ (0.0326) & $0.206$ (0.0371) & $0.107$ (0.0355) \\ 
$30$ & $30$ & $50$ & $200$ & $0.138$ (0.0395) & $0.091$ (0.0328) & $0.212$ (0.0389) & $0.109$ (0.0395) \\ 
$15$ & $30$ & $100$ & $200$ & $0.088$ (0.0255) & $0.091$ (0.0320) & $0.161$ (0.0291) & $0.077$ (0.0256) \\ 
$30$ & $30$ & $100$ & $200$ & $0.098$ (0.0280) & $0.090$ (0.0309) & $0.171$ (0.0295) & $0.081$ (0.0270) \\ 
$15$ & $30$ & $200$ & $200$ & $0.063$ (0.0184) & $0.090$ (0.0330) & $0.120$ (0.0209) & $0.057$ (0.0181) \\ 
$30$ & $30$ & $200$ & $200$ & $0.069$ (0.0200) & $0.091$ (0.0314) & $0.130$ (0.0224) & $0.060$ (0.0192) \\ \hline
\end{tabular} } }
\end{table}

\begin{table}[H] \centering 
  \caption{\com{Average empirical KS distances (standard deviations in brackets) for Kaplan--Meier estimator using only RC data (KM (RC)), the survival function NPMLE using only LBRC data (NPMLE (LBRC)), the survival function NPMLE using both RC and LBRC data (NPMLE RC+LBRC) and the density ratio model estimator (DRM). Failure times are gamma distributed with increasing hazard with varying censoring proportions and sample sizes for RC and LBRC data (RC \%/SS, LBRC \%/SS, respectively).}} 
  \label{corr_spec_inchaz}
  {   \renewcommand{\arraystretch}{1.2}   \resizebox{\textwidth}{!}{
\begin{tabular}{cccc|cccc} 
RC \% & LBRC \% & RC SS & LBRC SS & KM (RC) & NPMLE (LBRC) & NPMLE (RC+LBRC) & DRM \\ \hline
$15$ & $15$ & $50$ & $50$ & $0.126$ (0.0384) &  $0.156$ (0.0541) & $0.145$ (0.0422) & $0.112$ (0.0376) \\ 
$30$ & $15$ & $50$ & $50$ & $0.141$ (0.0424) &  $0.156$ (0.0531) & $0.152$ (0.0453) & $0.120$ (0.0418) \\ 
$15$ & $15$ & $100$ & $50$ & $0.091$ (0.0287) &  $0.156$ (0.0534) & $0.105$ (0.0316) & $0.084$ (0.0278) \\ 
$30$ & $15$ & $100$ & $50$ & $0.098$ (0.0295) &  $0.156$ (0.0540) & $0.113$ (0.0336) & $0.088$ (0.0287) \\ 
$15$ & $15$ & $200$ & $50$ & $0.065$ (0.0200) &  $0.154$ (0.0531) & $0.075$ (0.0241) & $0.062$ (0.0201) \\ 
$30$ & $15$ & $200$ & $50$ & $0.068$ (0.0199) &  $0.154$ (0.0553) & $0.079$ (0.0234) & $0.064$ (0.0198) \\ \hline \hline 
$15$ & $30$ & $50$ & $50$ & $0.128$ (0.0378) &  $0.162$ (0.0567) & $0.143$ (0.0428) & $0.114$ (0.0360) \\ 
$30$ & $30$ & $50$ & $50$ & $0.136$ (0.0399) &  $0.161$ (0.0558) & $0.153$ (0.0429) & $0.118$ (0.0388) \\ 
$15$ & $30$ & $100$ & $50$ & $0.088$ (0.0269) &  $0.159$ (0.0555) & $0.105$ (0.0310) & $0.082$ (0.0271) \\ 
$30$ & $30$ & $100$ & $50$ & $0.098$ (0.0305) &  $0.162$ (0.0555) & $0.114$ (0.0335) & $0.090$ (0.0298) \\ 
$15$ & $30$ & $200$ & $50$ & $0.064$ (0.0196) &  $0.158$ (0.0527) & $0.073$ (0.0232) & $0.062$ (0.0202) \\ 
$30$ & $30$ & $200$ & $50$ & $0.070$ (0.0201) &  $0.161$ (0.0545) & $0.080$ (0.0238)& $0.066$ (0.0204) \\  \hline \hline
$15$ & $15$ & $50$ & $100$ & $0.126$ (0.0387) &  $0.112$ (0.0390) & $0.167$ (0.0396) & $0.105$ (0.0371) \\ 
$30$ & $15$ & $50$ & $100$ & $0.137$ (0.0400) &  $0.110$ (0.0378) & $0.176$ (0.0405) & $0.110$ (0.0387) \\ 
$15$ & $15$ & $100$ & $100$ & $0.091$ (0.0278) &  $0.110$ (0.0384) & $0.126$ (0.0324) & $0.080$ (0.0272) \\ 
$30$ & $15$ & $100$ & $100$ & $0.098$ (0.0291) &  $0.112$ (0.0375) & $0.135$ (0.0329) & $0.083$ (0.0280) \\ 
$15$ & $15$ & $200$ & $100$ & $0.064$ (0.0191) &  $0.110$ (0.0349) & $0.091$ (0.0237) & $0.059$ (0.0187) \\ 
$30$ & $15$ & $200$ & $100$ & $0.068$ (0.0192) &  $0.112$ (0.0379) & $0.099$ (0.0249) & $0.061$ (0.0191) \\ \hline \hline 
$15$ & $30$ & $50$ & $100$ & $0.128$ (0.0393) &  $0.114$ (0.0368) & $0.166$ (0.0397) & $0.108$ (0.0384) \\ 
$30$ & $30$ & $50$ & $100$ & $0.135$ (0.0391) &  $0.115$ (0.0374) & $0.171$ (0.0411) & $0.109$ (0.0374) \\ 
$15$ & $30$ & $100$ & $100$ & $0.090$ (0.0282) &  $0.118$ (0.0390) & $0.127$ (0.0321)& $0.081$ (0.0275) \\ 
$30$ & $30$ & $100$ & $100$ & $0.097$ (0.0284) &  $0.117$ (0.0380) & $0.134$ (0.0328) & $0.084$ (0.0272) \\ 
$15$ & $30$ & $200$ & $100$ & $0.064$ (0.0190) &  $0.114$ (0.0380) & $0.089$ (0.0240) & $0.060$ (0.0192) \\ 
$30$ & $30$ & $200$ & $100$ & $0.070$ (0.0212) &  $0.117$ (0.0383) & $0.098$ (0.0258) & $0.064$ (0.0207) \\ \hline \hline 
$15$ & $15$ & $50$ & $200$ & $0.126$ (0.0396) &  $0.080$ (0.0262) & $0.194$ (0.0348) & $0.098$ (0.0365) \\ 
$30$ & $15$ & $50$ & $200$ & $0.138$ (0.0404) &  $0.079$ (0.0257) & $0.199$ (0.0341)  & $0.100$ (0.0362) \\ 
$15$ & $15$ & $100$ & $200$ & $0.090$ (0.0271) &  $0.079$ (0.0255) & $0.154$ (0.0287) & $0.075$ (0.0260) \\ 
$30$ & $15$ & $100$ & $200$ & $0.099$ (0.0296) &  $0.080$ (0.0256) & $0.161$ (0.0297) & $0.078$ (0.0271) \\ 
$15$ & $15$ & $200$ & $200$ & $0.065$ (0.0199) &  $0.079$ (0.0268) & $0.117$ (0.0231)  & $0.057$ (0.0192) \\ 
$30$ & $15$ & $200$ & $200$ & $0.070$ (0.0204) &  $0.080$ (0.0255) & $0.124$ (0.0238) & $0.059$ (0.0192) \\ \hline \hline 
$15$ & $30$ & $50$ & $200$ & $0.124$ (0.0379) &  $0.083$ (0.0273) & $0.193$ (0.0338)  & $0.096$ (0.0339) \\ 
$30$ & $30$ & $50$ & $200$ & $0.136$ (0.0410) &  $0.083$ (0.0270) & $0.197$ (0.0357) & $0.101$ (0.0382) \\ 
$15$ & $30$ & $100$ & $200$ & $0.088$ (0.0263) &  $0.082$ (0.0264) & $0.153$ (0.0276) & $0.073$ (0.0249) \\ 
$30$ & $30$ & $100$ & $200$ & $0.097$ (0.0294) &  $0.081$ (0.0263) & $0.161$ (0.0293)  & $0.078$ (0.0269) \\ 
$15$ & $30$ & $200$ & $200$ & $0.065$ (0.0199) &  $0.082$ (0.0258) & $0.115$ (0.0238)  & $0.058$ (0.0192) \\ 
$30$ & $30$ & $200$ & $200$ & $0.070$ (0.0212) &  $0.082$ (0.0277) & $0.122$ (0.0238) & $0.059$ (0.0202) \\ \hline
\end{tabular} } }
\end{table}

\newpage

\section*{Acknowledgments}

J.H. McVittie is supported by a Natural Sciences and Engineering Research Council of Canada Discovery Grant (RGPIN-2024-04763). 
A.G. Zhang gratefully acknowledges support from the Department of Statistical Sciences at the University of Toronto during his postdoctoral fellowship, where part of this work was completed.
The authors also thank the Digital Research Alliance of Canada for computing support. 

\bibliographystyle{apalike}
\bibliography{references}

\newpage

\setcounter{table}{0}
\renewcommand{\thetable}{S\arabic{table}}

\begin{center}
    {\LARGE \textbf{SUPPLEMENTARY MATERIALS}}
\end{center}
\vspace{1cm}

This supplementary material presents \com{some theoretical results on the proposed estimators,} additional simulation results evaluating the Kolmogorov--Smirnov distances between the proposed density ratio model (DRM) estimator and the truth under model overspecification, misspecification, and identical distribution settings \com{as well as additional plots assessing the model assumptions for the Montreal Hospital Duration application}. 

\com{\subsection*{Theoretical Results of EM Algorithm}}

\com{Although there is a close connection between the presented methodology and the biased sampling model literature, there are some subtle differences. In most analyses, it is typically assumed that the sampled biased failure times are fully observed and are not right-censored whereas here we assume the failure times drawn from both the reference and tilted densities could be censored. Following the work of \citetsupp{vardi}, we first list some numerical properties of our combined data estimator:}
\begin{enumerate}
    \item \com{There exists a maximizer, $\hat{p},\hat{\alpha},\hat{\theta}$, of the likelihood function for the set of combined data under the DRM assumption.}
    \item \com{The likelihood described in the previous item, increases with each iteration of the EM algorithm described in Section 2.2.}
    \item \com{The algorithm described in Section 2.2 converges to $\hat{p}, \hat{\alpha}, \hat{\theta}$.}
\end{enumerate}
\com{Unlike the setting of \citetsupp{vardi} with a single sample of length-biased right-censored failure time data, where he proved the NPMLE was unique, the same result does not hold in general when given multiple biased and possibly right-censored samples. \citetsupp{david} provided some technical conditions for checking for uniqueness when multiple biased samples were uncensored. The conditions to guarantee uniqueness of the NPMLE in the general setting remain open. The existence of the maximizer follows from the fact that the likelihood is log-concave but not strictly over a compact set. We appeal to Vardi's arguments to establish items 2 and 3. Since the proposed algorithm is an EM algorithm, it increases with each iteration. By the results of \citetsupp{csisz}, the EM algorithm converges to $\hat{p}, \hat{\alpha}, \hat{\theta}$}.

\com{\subsection*{Proof of Theorem 2.1}}

\com{The proof of Theorem 2.1 is based on the proof technique of \citetsupp{zhu2} for a single sample of left-truncated right-censored failure time data drawn from an exponentially tilted distribution. First, we remark that maximization of the likelihood given in Equation (1) with respect to $\alpha$, $\theta$ and $\mathrm{d}F_0$ under the constraint that $\alpha = \int_{0}^{\infty} e^{\theta h(x)}\mathrm{d}F_0(x)$ is equivalent to maximization of the likelihood given by:
$$L(\theta, \mathrm{d}F_0) = \left\{ \prod_{i=1}^{n_{\text{RC}}} \left[ \mathrm{d}F_0(x_i)\right]^{\delta_i} \left[1 - F_0(x_i)\right]^{1-\delta_i} \right\} \left\{ \prod_{i=1}^{n_{\text{LBRC}}} \frac{ \left[\exp\left\{\theta h(x_i')\right\} \mathrm{d}F_0(x_i')\right]^{\delta_i'} \left[1 - F_{\text{exptilt}}(x_i'; \theta, F_0)\right]^{1-\delta_i'}}{\mu(\theta, F_0)} \right\}$$
where $F_{\text{exptilt}}(x_i'; \theta, F_0) = \int_{x_i'}^{\infty} \exp(\theta h(u))\mathrm{d}F_0(u)$ and $\mu(\theta, F_0) = \int_{0}^{\infty} \int_{x_i'}^{\infty} \exp(\theta h(u)) \mathrm{d}F_0(u)$. Let $t_1 < \cdots < t_k$ denote the unique ordered failure times from the right-censored data set and the unique, ordered failure/censoring times from the length-biased right-censored data with probability masses $p_h = \mathrm{d}F_0(t_h)$ such that $\sum_{h=1}^{k} p_h = 1$. Let $n_h = \sum_{i=1}^{n_{\text{RC}}} I(X_i = t_h, \delta_i = 1)$, $n_h' = \sum_{i=1}^{n_{\text{LBRC}}} I(X_i' = t_h, \delta_i' = 1)$, $m_h = \sum_{i=1}^{n_{\text{RC}}} I(X_i = t_h, \delta_i = 0)$ and $m_h' = \sum_{i=1}^{n_{\text{LBRC}}} I(X_i' = t_h, \delta_i' = 0)$. Thus, using the discretized version of $\mathrm{d}F_0$ and the counts of the failure/censoring times at each mass point, the log-likelihood is given by:
$$\ell(\theta, p) = \sum_{h=1}^{k} n_h \log(p_h) + \sum_{h=1}^{k} m_h \log\left( \sum_{j=h}^{k} p_j\right) + \sum_{h=1}^{k} n_h'(\theta h(t_h)) + \sum_{h=1}^{k} n_h'\log(p_h) +$$
$$\sum_{h=1}^{k} m_h' \log\left[ \sum_{j=h}^{k} p_j \exp(\theta h(t_j))\right] - n_{\text{LBRC}} \log \left\{ \sum_{h=1}^{k} \left[ \sum_{j=h}^{k} p_j \exp(\theta h(t_h)) \right] \right\}.$$
Now, using this representation, following the approach of \citetsupp{zhu2}, we will reparameterize the above log-likelihood and use the same argument of \citetsupp{luo} that since the log-likelihood is linear in its parameters subject to a convex constraint space, that the maximizer exists and is necessarily unique. Let $g_h = \frac{1}{p_h} = \exp(\gamma_h)$, $u_h = \sum_{j=h}^{k} \left(\frac{1}{g_h}\right)\exp(\theta h(t_j)) = \exp(\mu_h)$, $v_h = \sum_{j=h}^{k} \left(\frac{1}{g_h}\right) = \exp(\nu_h)$ and $\rho = \sum_{h=1}^{k} u_h = \exp(\sigma)$. Thus maximization is $\ell(\theta, p)$ is equivalent to minimization of 
$$s(\gamma, \mu, \nu, \sigma) = \sum_{h=1}^{k} n_h \gamma_h - \sum_{h=1}^{k} m_h \nu_h - \sum_{h=1}^{k} n_h'(\theta h(t_h)) - \sum_{h=1}^{k} n_h'\gamma_h - \sum_{h=1}^{H} m_h'\mu_h + n_{\text{LBRC}} \sigma,$$
subject to the constraints $\sum_{h=1}^{k} \frac{1}{g_h} = 1$, $\sum_{j=1}^{k} \exp(\theta h(y_j))\left(\frac{1}{g_h}\right) = \exp(\mu_h) = u_h$, $\sum_{h=1}^{k} u_h = \exp(\sigma)$, $v_h = \sum_{j=h}^{k} \frac{1}{g_j} = \exp(\nu_h)$ and $g_h > 0$ for all $h = 1, \ldots, k$. Following the argument of \citetsupp{luo}, this minimization is equivalent to minimization of $s$ subject to the constraints $\sum_{h=1}^{k} \exp(-\gamma_h) \leq 1$, $\sum_{j=h}^{k} \exp(\theta h(t_j))\exp(-(\gamma_j + \mu_h)) \leq 1$, $\sum_{j=h}^{k} \exp(-(\gamma_j + \nu_h)) \leq 1$ and $\sum_{h=1}^{k} \exp(\mu_h - \sigma) \leq 1$. As argued in \citetsupp{zhu2}, since the constraint region is convex and since the function $s(\gamma, \mu, \nu, \sigma)$ is linear in its parameters, there exists a unique minimizer of $s(\gamma, \mu, \nu, \sigma)$ and thus a unique maximizer of $\ell(\theta, p)$.}

\com{To establish that the estimators $\hat{\theta}$ and $\hat{F}_0$ are consistent, we use the proof approach of \citetsupp{murph} as applied by \citetsupp{mcvit2} based on a Kullback--Leibler divergence argument. Let $\xi = (\theta, F_0)$. Since $\hat{\xi}_n$ is bounded then by Helly's selection theorem, there exists a convergent subsequence $\hat{\xi}_{n_k}$ such that $\hat{\theta}_{n_k} \rightarrow \theta^*$ and $\hat{F}_{0, n_k} \rightarrow F_0^*$ for some $\theta^*$ and $F_{0}^{*}$. Thus, the maximum likelihood estimators $\hat{\theta}$ and $\hat{F}_0$ satisfy:
$$\ell(\hat{\theta}, \hat{F}_0) - \ell(\bar{\theta}, \bar{F}_0) \geq 0$$
for all $\bar{\theta}$ and $\bar{F}_0$. Assume that the ratio of the right-censored data to the combined data given by:
$$\hat{\beta} = \frac{n_{\text{RC}}}{n_{\text{RC}} + n_{\text{LBRC}}} \rightarrow \beta \in (0, 1).$$
Now, consider the normalized log-likelihood given by:
$$\frac{1}{n_{\text{RC}} + n_{\text{LBRC}}} l(\theta, F_0) = \hat{\beta} \frac{1}{n_{\text{RC}}}\ell_{{\text{RC}}} (\theta, F_0) + (1-\hat{\beta}) \frac{1}{n_{\text{LBRC}}}\ell_{{\text{LBRC}}}(\theta, F_0),$$
where
$$\ell_{{\text{RC}}}(\theta, F_0) = \sum_{i=1}^{n_{\text{RC}}} \left[ \delta_i \log \mathrm{d}F_0(X_i) + (1-\delta_i) \log \left\{ \int_{X_i}^{\infty} \mathrm{d}F_0(t) \right\}\right]$$
and 
$$\ell_{{\text{LBRC}}}(\theta, F_0) = \sum_{i=1}^{n_{\text{LBRC}}} \bigg[ \delta_i' \log \mathrm{d}F_0(X_i') + (1-\delta_i') \log \left\{ \int_{X_i'}^{\infty} \mathrm{d}F_0(t) \right\}-$$
$$ \log\left\{ \int_{0}^{\infty} \left[ \int_{t}^{\infty} \exp(\theta h(u)) \mathrm{d}F_0(u) \right] \right\} \bigg].$$
Now, from the above inequality, we have that  
$$\ell(\hat{\theta}_{n_k}, \hat{F}_{0,n_k}) - \ell(\theta_0, F_0) \geq 0,$$
for the true $\theta_0$ and $F_0$ parameters. Observe that as $n_{\text{RC}} \rightarrow \infty$ and $n_{\text{LBRC}} \rightarrow \infty$, by the strong law of large numbers, the difference
$$\frac{1}{n_{\text{RC}, k} + n_{\text{LBRC}, k}}\left( \ell(\hat{\theta}_{n_k}, \hat{F}_{0,n_k}) - \ell(\theta_0, F_0)\right) \rightarrow \beta  \mathbb{E}_0\left[\ell_{\text{RC}}(\theta^*, F_0^*) - \ell_{\text{RC}}(\theta_0, F_0)\right]  +$$
$$(1-\beta) \mathbb{E}_0\left[\ell_{\text{LBRC}}(\theta^*, F_0^*) - \ell_{\text{LBRC}}(\theta_0, F_0) \right].$$
But we note that each of these terms represent the negative Kullback--Leibler divergence and so the linear combination will also be negative. This implies that $\mathbb{E}_0\left(\ell(\theta^*, F_0^*)\right) = \mathbb{E}_0\left(\ell(\theta_0, F_0)\right)$ which by parameter identifiability, implies that $\theta^* = \theta_0$ for any fixed $t \in [0, \tau]$. Applying the arguments of \citetsupp{murph}, we obtain $\hat{\theta}\rightarrow \theta_0$ and for fixed $t \in [0, \tau]$, $\hat{F}_0(t) \rightarrow F_0(t)$. Since $\hat{F}_0(t)$ are monotone functions and since $F_0(t)$ is monotone, then uniform convergence of $\hat{F}_0(t)$ follows from \citetsupp{doob}. \qed}

\com{One approach to show that the maximum likelihood estimators are asymptotically normally distributed is through the use of empirical process theory as described in \citetsupp{vaart}. While these proof techniques were applied by \citetsupp{mcvit2} in the case where $\alpha = \theta = 0$ to show $\hat{F}_0$ was weakly convergent, they required unverifiable assumptions on the censoring distributions and cumulative hazard function. Furthermore, the asymptotic covariance matrices did not yield closed-form expressions that could be directly evaluated, ultimately requiring the use of the nonparametric bootstrapping procedure. Due to these theoretical limitations on establishing the weak convergence of the tilting parameters and reference distribution, we do not provide any details on the proof.}

\com{\subsection*{Additional Simulation Results}}

In Scenarios~(i) and~(ii), we simulate right-censored (RC) and length-biased right-censored (LBRC) failure times using the same gamma distribution settings as in the main text. 
In Scenario~(iii), both the RC and LBRC failure times are generated from an identical gamma distribution with shape parameter $0.5$, $1$, or $2$ and scale parameter $2$. 

To make the results self-contained, we also include the performance of the competing methods considered in the main text in all tables, to facilitate better comparison for readers.

\begin{enumerate}
\item[(i)] Tables~\ref{supp_overspec_dechaz}--\ref{supp_overspec_inchaz} summarize the simulation results when the DRM is overspecified with the vector-valued basis functions $h(x)=(x, \log(x))$, when the most suitable basis function is $h^*(x)=\log(x)$, corresponding to the decreasing-, constant-, and increasing-hazard settings, respectively.
Such an overspecified model covers the gamma family without the equal-scale-parameter assumption.

\item[(ii)] Tables~\ref{supp_misspec1_dechaz}--\ref{supp_misspec1_inchaz} provide the simulation results when the DRM is misspecified with the basis functions $h(x)=\sqrt{x}$, corresponding to the decreasing-, constant-, and increasing-hazard settings, respectively.

\noindent Similarly, Tables~\ref{supp_misspec2_dechaz}--\ref{supp_misspec2_inchaz} show the results when the DRM is misspecified with the basis functions $h(x)=x$.

\item[(iii)] Tables~\ref{supp_identical_dechaz}--\ref{supp_identical_inchaz} present the simulation results for the case where the RC and LBRC failure time distributions are identical, both from a gamma distribution with shape parameter $0.5$, $1$, or $2$ and scale parameter $2$, while the DRM is still fitted using the basis function $h(x)=\log(x)$ without assuming distributional equivalence.
To examine whether a larger sample size improves the DRM’s ability to recognize distributional similarity, we increase the sample sizes by a factor of ten relative to those in the main text. 
The three tables correspond to the decreasing-, constant-, and increasing-hazard settings, respectively.
\end{enumerate}

\com{\subsection*{Applications - Model Assumptions}}

\begin{figure}
\centering
    \includegraphics{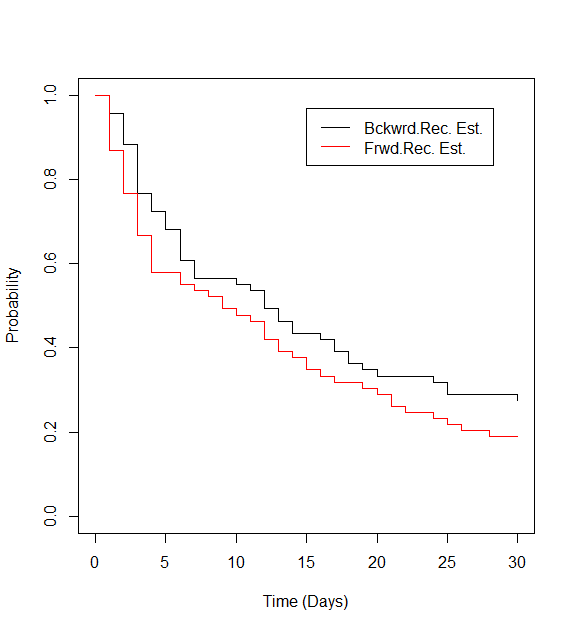}
    \caption{\com{Survival curve estimates of the observed forward and backward recurrence times for the Montreal hospital length of stay durations.}}
    \label{bckfwrcompar_plot}
\end{figure}

\begin{figure}
\centering
    \includegraphics{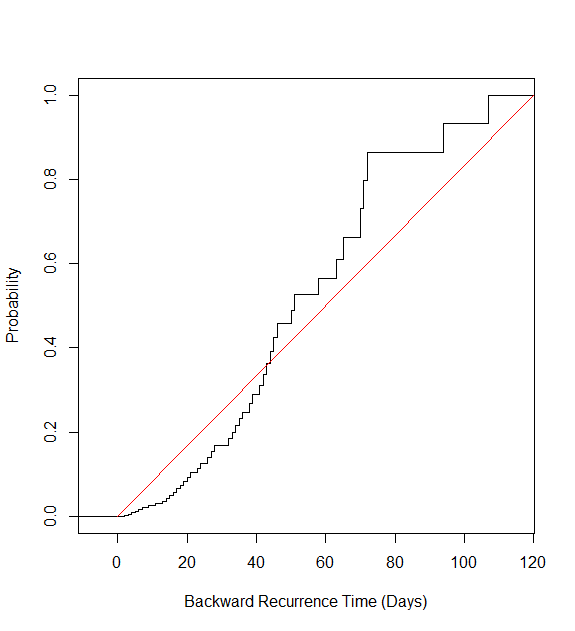}
    \caption{\com{Nonparametric cumulative distribution function estimate of the underlying left-truncation distribution (black) for the Montreal hospital length of stay durations compared to the cumulative distribution function of a Uniform distribution over $(0,120)$.}}
    \label{bcktrue_plot}
\end{figure}

\begin{figure}
\centering
    \includegraphics{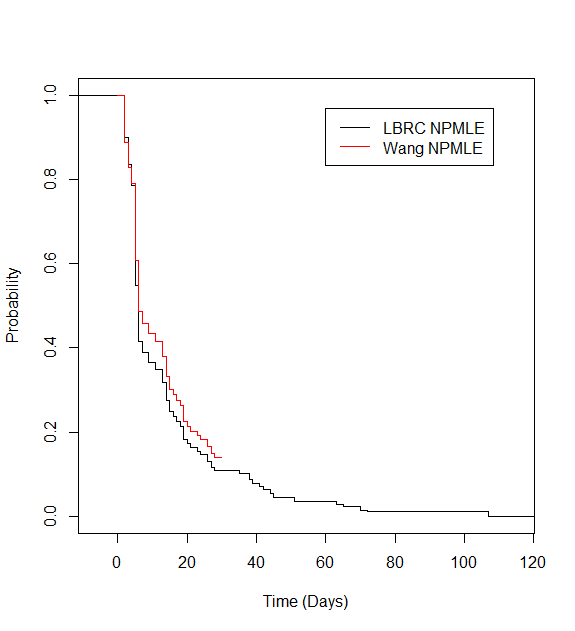}
    \caption{\com{Comparison of nonparametric survival function estimates for the Montreal hospital length of stay durations using the robust estimator of Wang (black) and the survival function NPMLE under the stationarity assumption (red).}}
    \label{survcompar_plot}
\end{figure}

\com{We assessed the stationarity assumption for the left-truncated right-censored hospital stay durations graphically in three different ways as proposed by \citetsupp{asgha3}. First, under the stationarity assumption, it can be shown that the distributions of the backward recurrence times and the forward recurrence times are equivalent. Using this equivalence, we fit the survival function Kaplan--Meier estimates for each and compared their relative shapes in Figure \ref{bckfwrcompar_plot}. From the estimated curves for the forward and backward recurrence times, they generally have the same shape although due to the limited follow-up in the forward recurrence times, the right tail of the forward recurrence time estimate was undefined past 30 days. We also estimated the underlying left-truncation cumulative distribution function using the nonparametric estimator proposed by \citetsupp{wang2} and compared it to the uniform cumulative distribution function in Figure \ref{bcktrue_plot}. We note that we removed an outlier of $200$ days when fitting the curve as the majority of the failure/censoring times were found between $2$ and $103$ days. The figure shows that the nonparametric estimate generally follows the linear shape of a uniform cumulative distribution function. Finally, we compared the survival function NPMLE under a general truncation mechanism, as proposed by \citetsupp{wang2}, to the survival function NPMLE under the stationarity assumption in Figure \ref{survcompar_plot}. If the stationarity assumption is valid, the two curves should have roughly the same shape. From the estimated curves, indeed, we find that the survival function estimates are relatively similar, indicating that the stationarity assumption appears valid for the Montreal hospital stay duration data.}

\vspace{1cm}
\bibliographystylesupp{apalike}
\bibliographysupp{references}

\newpage

\begin{table}[!htbp] \centering 
  \caption{Comparison of average empirical Kolmogorov--Smirnov distances computed over $1000$ simulation replications for gamma distributed failure times with decreasing hazard, varying censoring proportions and sample sizes for the RC and LBRC data (RC \%/SS, LBRC \%/SS, respectively). Estimates include the Kaplan--Meier estimator using only the RC data (KM (RC)), the adjusted Kaplan--Meier estimator using only the LBRC data while ignoring length-bias structure (KM (LTRC)), the survival function NPMLE using only the LBRC data (NPMLE (LBRC)), the survival function NPMLE using both RC and LBRC data (NPMLE RC+LBRC) and the overspecified density ratio model estimator (DRM) with $h(x) = (x, \log(x))$.} 
  \label{supp_overspec_dechaz} 
  {   \renewcommand{\arraystretch}{1.3}   \resizebox{\textwidth}{!}{
\begin{tabular}{cccc|ccccc}
RC \% & LBRC \% & RC SS & LBRC SS & KM (RC) & KM (LTRC) & NPMLE (LBRC) & NPMLE (RC+LBRC) & DRM \\ \hline

$15$ & $15$ & $50$ & $50$ & $0.124$ & $0.209$ & $0.193$ & $0.140$ & $0.116$ \\ 
$30$ & $15$ & $50$ & $50$ & $0.145$ & $0.214$ & $0.191$ & $0.150$ & $0.127$ \\ 
$15$ & $15$ & $100$ & $50$ & $0.088$ & $0.210$ & $0.193$ & $0.103$ & $0.085$ \\ 
$30$ & $15$ & $100$ & $50$ & $0.102$ & $0.210$ & $0.192$ & $0.116$ & $0.091$ \\ 
$15$ & $15$ & $200$ & $50$ & $0.062$ & $0.213$ & $0.192$ & $0.074$ & $0.060$ \\ 
$30$ & $15$ & $200$ & $50$ & $0.074$ & $0.212$ & $0.193$ & $0.085$ & $0.066$ \\ 
$15$ & $30$ & $50$ & $50$ & $0.125$ & $0.219$ & $0.199$ & $0.142$ & $0.118$ \\ 
$30$ & $30$ & $50$ & $50$ & $0.146$ & $0.218$ & $0.197$ & $0.153$ & $0.130$ \\ 
$15$ & $30$ & $100$ & $50$ & $0.088$ & $0.218$ & $0.198$ & $0.106$ & $0.084$ \\ 
$30$ & $30$ & $100$ & $50$ & $0.100$ & $0.218$ & $0.196$ & $0.113$ & $0.090$ \\ 
$15$ & $30$ & $200$ & $50$ & $0.062$ & $0.221$ & $0.197$ & $0.073$ & $0.060$ \\ 
$30$ & $30$ & $200$ & $50$ & $0.074$ & $0.218$ & $0.196$ & $0.085$ & $0.067$ \\ 
$15$ & $15$ & $50$ & $100$ & $0.122$ & $0.155$ & $0.143$ & $0.164$ & $0.113$ \\ 
$30$ & $15$ & $50$ & $100$ & $0.143$ & $0.154$ & $0.141$ & $0.173$ & $0.121$ \\ 
$15$ & $15$ & $100$ & $100$ & $0.087$ & $0.157$ & $0.142$ & $0.124$ & $0.082$ \\ 
$30$ & $15$ & $100$ & $100$ & $0.103$ & $0.154$ & $0.141$ & $0.135$ & $0.088$ \\ 
$15$ & $15$ & $200$ & $100$ & $0.062$ & $0.156$ & $0.141$ & $0.090$ & $0.060$ \\ 
$30$ & $15$ & $200$ & $100$ & $0.073$ & $0.154$ & $0.142$ & $0.102$ & $0.064$ \\ 
$15$ & $30$ & $50$ & $100$ & $0.125$ & $0.160$ & $0.147$ & $0.164$ & $0.116$ \\ 
$30$ & $30$ & $50$ & $100$ & $0.144$ & $0.161$ & $0.146$ & $0.172$ & $0.124$ \\ 
$15$ & $30$ & $100$ & $100$ & $0.088$ & $0.159$ & $0.146$ & $0.124$ & $0.083$ \\ 
$30$ & $30$ & $100$ & $100$ & $0.101$ & $0.162$ & $0.146$ & $0.134$ & $0.088$ \\ 
$15$ & $30$ & $200$ & $100$ & $0.061$ & $0.157$ & $0.144$ & $0.088$ & $0.059$ \\ 
$30$ & $30$ & $200$ & $100$ & $0.073$ & $0.159$ & $0.146$ & $0.101$ & $0.064$ \\ 
$15$ & $15$ & $50$ & $200$ & $0.124$ & $0.114$ & $0.105$ & $0.193$ & $0.112$ \\ 
$30$ & $15$ & $50$ & $200$ & $0.144$ & $0.114$ & $0.105$ & $0.203$ & $0.117$ \\ 
$15$ & $15$ & $100$ & $200$ & $0.087$ & $0.113$ & $0.105$ & $0.151$ & $0.079$ \\ 
$30$ & $15$ & $100$ & $200$ & $0.102$ & $0.115$ & $0.107$ & $0.163$ & $0.085$ \\ 
$15$ & $15$ & $200$ & $200$ & $0.062$ & $0.111$ & $0.104$ & $0.114$ & $0.058$ \\ 
$30$ & $15$ & $200$ & $200$ & $0.073$ & $0.109$ & $0.104$ & $0.126$ & $0.062$ \\ 
$15$ & $30$ & $50$ & $200$ & $0.123$ & $0.113$ & $0.106$ & $0.194$ & $0.110$ \\ 
$30$ & $30$ & $50$ & $200$ & $0.142$ & $0.117$ & $0.107$ & $0.204$ & $0.116$ \\ 
$15$ & $30$ & $100$ & $200$ & $0.089$ & $0.117$ & $0.108$ & $0.151$ & $0.082$ \\ 
$30$ & $30$ & $100$ & $200$ & $0.103$ & $0.113$ & $0.104$ & $0.162$ & $0.087$ \\ 
$15$ & $30$ & $200$ & $200$ & $0.062$ & $0.112$ & $0.105$ & $0.113$ & $0.058$ \\ 
$30$ & $30$ & $200$ & $200$ & $0.073$ & $0.113$ & $0.104$ & $0.126$ & $0.062$ \\ 
\hline \\[-1.8ex] 
\end{tabular} } }
\end{table} 

\begin{table}[!htbp] \centering 
  \caption{Comparison of average empirical Kolmogorov--Smirnov distances computed over $1000$ simulation replications for gamma distributed failure times with constant hazard, varying censoring proportions and sample sizes for the RC and LBRC data (RC \%/SS, LBRC \%/SS, respectively). Estimates include the Kaplan--Meier estimator using only the RC data (KM (RC)), the adjusted Kaplan--Meier estimator using only the LBRC data while ignoring length-bias structure (KM (LTRC)), the survival function NPMLE using only the LBRC data (NPMLE (LBRC)), the survival function NPMLE using both RC and LBRC data (NPMLE RC+LBRC) and the overspecified density ratio model estimator (DRM) with $h(x) = (x, \log(x))$.} 
  \label{supp_overspec_consthaz} 
  {   \renewcommand{\arraystretch}{1.3}   \resizebox{\textwidth}{!}{
\begin{tabular}{cccc|ccccc}
RC \% & LBRC \% & RC SS & LBRC SS & KM (RC) & KM (LTRC) & NPMLE (LBRC) & NPMLE (RC+LBRC) & DRM \\ \hline
$15$ & $15$ & $50$ & $50$ & $0.125$ & $0.205$ & $0.185$ & $0.149$ & $0.116$ \\ 
$30$ & $15$ & $50$ & $50$ & $0.139$ & $0.200$ & $0.183$ & $0.158$ & $0.125$ \\ 
$15$ & $15$ & $100$ & $50$ & $0.089$ & $0.196$ & $0.182$ & $0.107$ & $0.084$ \\ 
$30$ & $15$ & $100$ & $50$ & $0.097$ & $0.197$ & $0.181$ & $0.116$ & $0.090$ \\ 
$15$ & $15$ & $200$ & $50$ & $0.063$ & $0.196$ & $0.180$ & $0.074$ & $0.061$ \\ 
$30$ & $15$ & $200$ & $50$ & $0.070$ & $0.195$ & $0.179$ & $0.083$ & $0.066$ \\ 
$15$ & $30$ & $50$ & $50$ & $0.127$ & $0.187$ & $0.171$ & $0.148$ & $0.116$ \\ 
$30$ & $30$ & $50$ & $50$ & $0.139$ & $0.184$ & $0.172$ & $0.158$ & $0.124$ \\ 
$15$ & $30$ & $100$ & $50$ & $0.089$ & $0.194$ & $0.177$ & $0.107$ & $0.085$ \\ 
$30$ & $30$ & $100$ & $50$ & $0.099$ & $0.187$ & $0.173$ & $0.116$ & $0.090$ \\ 
$15$ & $30$ & $200$ & $50$ & $0.062$ & $0.190$ & $0.174$ & $0.074$ & $0.060$ \\ 
$30$ & $30$ & $200$ & $50$ & $0.069$ & $0.190$ & $0.173$ & $0.085$ & $0.065$ \\ 
$15$ & $15$ & $50$ & $100$ & $0.124$ & $0.143$ & $0.130$ & $0.171$ & $0.112$ \\ 
$30$ & $15$ & $50$ & $100$ & $0.138$ & $0.142$ & $0.130$ & $0.182$ & $0.119$ \\ 
$15$ & $15$ & $100$ & $100$ & $0.090$ & $0.142$ & $0.132$ & $0.129$ & $0.083$ \\ 
$30$ & $15$ & $100$ & $100$ & $0.097$ & $0.139$ & $0.128$ & $0.138$ & $0.086$ \\ 
$15$ & $15$ & $200$ & $100$ & $0.063$ & $0.139$ & $0.129$ & $0.093$ & $0.060$ \\ 
$30$ & $15$ & $200$ & $100$ & $0.068$ & $0.142$ & $0.130$ & $0.101$ & $0.063$ \\ 
$15$ & $30$ & $50$ & $100$ & $0.125$ & $0.139$ & $0.128$ & $0.171$ & $0.111$ \\ 
$30$ & $30$ & $50$ & $100$ & $0.138$ & $0.137$ & $0.127$ & $0.183$ & $0.116$ \\ 
$15$ & $30$ & $100$ & $100$ & $0.089$ & $0.135$ & $0.124$ & $0.130$ & $0.081$ \\ 
$30$ & $30$ & $100$ & $100$ & $0.098$ & $0.135$ & $0.127$ & $0.139$ & $0.086$ \\ 
$15$ & $30$ & $200$ & $100$ & $0.064$ & $0.140$ & $0.129$ & $0.094$ & $0.060$ \\ 
$30$ & $30$ & $200$ & $100$ & $0.068$ & $0.136$ & $0.125$ & $0.103$ & $0.063$ \\ 
$15$ & $15$ & $50$ & $200$ & $0.125$ & $0.101$ & $0.096$ & $0.208$ & $0.106$ \\ 
$30$ & $15$ & $50$ & $200$ & $0.138$ & $0.100$ & $0.094$ & $0.212$ & $0.111$ \\ 
$15$ & $15$ & $100$ & $200$ & $0.089$ & $0.102$ & $0.094$ & $0.160$ & $0.079$ \\ 
$30$ & $15$ & $100$ & $200$ & $0.098$ & $0.101$ & $0.094$ & $0.169$ & $0.084$ \\ 
$15$ & $15$ & $200$ & $200$ & $0.062$ & $0.099$ & $0.092$ & $0.119$ & $0.058$ \\ 
$30$ & $15$ & $200$ & $200$ & $0.069$ & $0.100$ & $0.094$ & $0.128$ & $0.061$ \\ 
$15$ & $30$ & $50$ & $200$ & $0.127$ & $0.098$ & $0.092$ & $0.206$ & $0.109$ \\ 
$30$ & $30$ & $50$ & $200$ & $0.138$ & $0.097$ & $0.090$ & $0.213$ & $0.111$ \\ 
$15$ & $30$ & $100$ & $200$ & $0.088$ & $0.095$ & $0.091$ & $0.161$ & $0.078$ \\ 
$30$ & $30$ & $100$ & $200$ & $0.098$ & $0.098$ & $0.090$ & $0.171$ & $0.083$ \\ 
$15$ & $30$ & $200$ & $200$ & $0.062$ & $0.097$ & $0.090$ & $0.120$ & $0.057$ \\ 
$30$ & $30$ & $200$ & $200$ & $0.069$ & $0.097$ & $0.091$ & $0.130$ & $0.060$ \\ 
\hline \\[-1.8ex] 
\end{tabular} } }
\end{table} 

\begin{table}[!htbp] \centering 
  \caption{Comparison of average empirical Kolmogorov--Smirnov distances computed over $1000$ simulation replications for gamma distributed failure times with increasing hazard, varying censoring proportions and sample sizes for the RC and LBRC data (RC \%/SS, LBRC \%/SS, respectively). Estimates include the Kaplan--Meier estimator using only the RC data (KM (RC)), the adjusted Kaplan--Meier estimator using only the LBRC data while ignoring length-bias structure (KM (LTRC)), the survival function NPMLE using only the LBRC data (NPMLE (LBRC)), the survival function NPMLE using both RC and LBRC data (NPMLE RC+LBRC) and the overspecified density ratio model estimator (DRM) with $h(x) = (x, \log(x))$.} 
  \label{supp_overspec_inchaz} 
  {   \renewcommand{\arraystretch}{1.3}   \resizebox{\textwidth}{!}{
\begin{tabular}{cccc|ccccc}
RC \% & LBRC \% & RC SS & LBRC SS & KM (RC) & KM (LTRC) & NPMLE (LBRC) & NPMLE (RC+LBRC) & DRM \\ \hline 
$15$ & $15$ & $50$ & $50$ & $0.126$ & $0.169$ & $0.157$ & $0.144$ & $0.113$ \\ 
$30$ & $15$ & $50$ & $50$ & $0.141$ & $0.167$ & $0.156$ & $0.152$ & $0.122$ \\ 
$15$ & $15$ & $100$ & $50$ & $0.091$ & $0.164$ & $0.154$ & $0.105$ & $0.085$ \\ 
$30$ & $15$ & $100$ & $50$ & $0.098$ & $0.168$ & $0.158$ & $0.113$ & $0.089$ \\ 
$15$ & $15$ & $200$ & $50$ & $0.065$ & $0.167$ & $0.155$ & $0.075$ & $0.062$ \\ 
$30$ & $15$ & $200$ & $50$ & $0.069$ & $0.168$ & $0.155$ & $0.080$ & $0.065$ \\ 
$15$ & $30$ & $50$ & $50$ & $0.128$ & $0.172$ & $0.162$ & $0.142$ & $0.115$ \\ 
$30$ & $30$ & $50$ & $50$ & $0.137$ & $0.175$ & $0.160$ & $0.152$ & $0.120$ \\ 
$15$ & $30$ & $100$ & $50$ & $0.088$ & $0.173$ & $0.160$ & $0.104$ & $0.083$ \\ 
$30$ & $30$ & $100$ & $50$ & $0.098$ & $0.174$ & $0.161$ & $0.113$ & $0.090$ \\ 
$15$ & $30$ & $200$ & $50$ & $0.064$ & $0.172$ & $0.159$ & $0.073$ & $0.062$ \\ 
$30$ & $30$ & $200$ & $50$ & $0.070$ & $0.175$ & $0.162$ & $0.080$ & $0.066$ \\ 
$15$ & $15$ & $50$ & $100$ & $0.126$ & $0.117$ & $0.112$ & $0.167$ & $0.106$ \\ 
$30$ & $15$ & $50$ & $100$ & $0.136$ & $0.121$ & $0.112$ & $0.176$ & $0.112$ \\ 
$15$ & $15$ & $100$ & $100$ & $0.091$ & $0.118$ & $0.111$ & $0.126$ & $0.082$ \\ 
$30$ & $15$ & $100$ & $100$ & $0.098$ & $0.118$ & $0.111$ & $0.134$ & $0.085$ \\ 
$15$ & $15$ & $200$ & $100$ & $0.064$ & $0.118$ & $0.111$ & $0.090$ & $0.060$ \\ 
$30$ & $15$ & $200$ & $100$ & $0.068$ & $0.119$ & $0.111$ & $0.099$ & $0.062$ \\ 
$15$ & $30$ & $50$ & $100$ & $0.127$ & $0.122$ & $0.115$ & $0.166$ & $0.109$ \\ 
$30$ & $30$ & $50$ & $100$ & $0.135$ & $0.124$ & $0.115$ & $0.170$ & $0.112$ \\ 
$15$ & $30$ & $100$ & $100$ & $0.090$ & $0.126$ & $0.117$ & $0.126$ & $0.081$ \\ 
$30$ & $30$ & $100$ & $100$ & $0.097$ & $0.127$ & $0.118$ & $0.133$ & $0.086$ \\ 
$15$ & $30$ & $200$ & $100$ & $0.064$ & $0.120$ & $0.113$ & $0.088$ & $0.060$ \\ 
$30$ & $30$ & $200$ & $100$ & $0.070$ & $0.124$ & $0.116$ & $0.098$ & $0.064$ \\ 
$15$ & $15$ & $50$ & $200$ & $0.127$ & $0.084$ & $0.079$ & $0.192$ & $0.102$ \\ 
$30$ & $15$ & $50$ & $200$ & $0.137$ & $0.085$ & $0.080$ & $0.199$ & $0.103$ \\ 
$15$ & $15$ & $100$ & $200$ & $0.090$ & $0.083$ & $0.078$ & $0.155$ & $0.076$ \\ 
$30$ & $15$ & $100$ & $200$ & $0.098$ & $0.085$ & $0.080$ & $0.161$ & $0.080$ \\ 
$15$ & $15$ & $200$ & $200$ & $0.064$ & $0.085$ & $0.080$ & $0.117$ & $0.057$ \\ 
$30$ & $15$ & $200$ & $200$ & $0.070$ & $0.085$ & $0.081$ & $0.124$ & $0.060$ \\ 
$15$ & $30$ & $50$ & $200$ & $0.124$ & $0.088$ & $0.083$ & $0.192$ & $0.099$ \\ 
$30$ & $30$ & $50$ & $200$ & $0.138$ & $0.089$ & $0.083$ & $0.196$ & $0.106$ \\ 
$15$ & $30$ & $100$ & $200$ & $0.089$ & $0.087$ & $0.082$ & $0.152$ & $0.076$ \\ 
$30$ & $30$ & $100$ & $200$ & $0.097$ & $0.086$ & $0.081$ & $0.161$ & $0.080$ \\ 
$15$ & $30$ & $200$ & $200$ & $0.064$ & $0.088$ & $0.083$ & $0.115$ & $0.058$ \\ 
$30$ & $30$ & $200$ & $200$ & $0.070$ & $0.087$ & $0.081$ & $0.122$ & $0.060$ \\ 
\hline \\[-1.8ex] 
\end{tabular} } }
\end{table}

\begin{table}[!htbp] \centering 
  \caption{Comparison of average empirical Kolmogorov--Smirnov distances computed over $1000$ simulation replications for gamma distributed failure times with decreasing hazard, varying censoring proportions and sample sizes for the RC and LBRC data (RC \%/SS, LBRC \%/SS, respectively). Estimates include the Kaplan--Meier estimator using only the RC data (KM (RC)), the adjusted Kaplan--Meier estimator using only the LBRC data while ignoring length-bias structure (KM (LTRC)), the survival function NPMLE using only the LBRC data (NPMLE (LBRC)), the survival function NPMLE using both RC and LBRC data (NPMLE RC+LBRC) and the misspecified density ratio model estimator (DRM) with $h(x) = \sqrt{x}$.} 
  \label{supp_misspec1_dechaz} 
  {   \renewcommand{\arraystretch}{1.3}   \resizebox{\textwidth}{!}{
\begin{tabular}{cccc|ccccc}
RC \% & LBRC \% & RC SS & LBRC SS & KM (RC) & KM (LTRC) & NPMLE (LBRC) & NPMLE (RC+LBRC) & DRM \\ \hline
 
$15$ & $15$ & $50$ & $50$ & $0.123$ & $0.209$ & $0.193$ & $0.140$ & $0.116$ \\ 
$30$ & $15$ & $50$ & $50$ & $0.145$ & $0.214$ & $0.191$ & $0.151$ & $0.124$ \\ 
$15$ & $15$ & $100$ & $50$ & $0.088$ & $0.210$ & $0.193$ & $0.103$ & $0.084$ \\ 
$30$ & $15$ & $100$ & $50$ & $0.102$ & $0.210$ & $0.192$ & $0.115$ & $0.090$ \\ 
$15$ & $15$ & $200$ & $50$ & $0.062$ & $0.213$ & $0.192$ & $0.074$ & $0.060$ \\ 
$30$ & $15$ & $200$ & $50$ & $0.074$ & $0.212$ & $0.192$ & $0.085$ & $0.066$ \\ 
$15$ & $30$ & $50$ & $50$ & $0.125$ & $0.219$ & $0.199$ & $0.142$ & $0.118$ \\ 
$30$ & $30$ & $50$ & $50$ & $0.145$ & $0.218$ & $0.197$ & $0.153$ & $0.125$ \\ 
$15$ & $30$ & $100$ & $50$ & $0.088$ & $0.218$ & $0.198$ & $0.106$ & $0.084$ \\ 
$30$ & $30$ & $100$ & $50$ & $0.100$ & $0.218$ & $0.196$ & $0.113$ & $0.088$ \\ 
$15$ & $30$ & $200$ & $50$ & $0.062$ & $0.221$ & $0.197$ & $0.073$ & $0.060$ \\ 
$30$ & $30$ & $200$ & $50$ & $0.073$ & $0.218$ & $0.195$ & $0.085$ & $0.066$ \\ 
$15$ & $15$ & $50$ & $100$ & $0.122$ & $0.154$ & $0.143$ & $0.164$ & $0.113$ \\ 
$30$ & $15$ & $50$ & $100$ & $0.143$ & $0.155$ & $0.141$ & $0.173$ & $0.118$ \\ 
$15$ & $15$ & $100$ & $100$ & $0.087$ & $0.156$ & $0.142$ & $0.124$ & $0.082$ \\ 
$30$ & $15$ & $100$ & $100$ & $0.103$ & $0.154$ & $0.141$ & $0.135$ & $0.087$ \\ 
$15$ & $15$ & $200$ & $100$ & $0.062$ & $0.156$ & $0.141$ & $0.090$ & $0.059$ \\ 
$30$ & $15$ & $200$ & $100$ & $0.073$ & $0.153$ & $0.142$ & $0.102$ & $0.063$ \\ 
$15$ & $30$ & $50$ & $100$ & $0.125$ & $0.160$ & $0.147$ & $0.164$ & $0.116$ \\ 
$30$ & $30$ & $50$ & $100$ & $0.144$ & $0.161$ & $0.146$ & $0.172$ & $0.121$ \\ 
$15$ & $30$ & $100$ & $100$ & $0.088$ & $0.159$ & $0.146$ & $0.124$ & $0.083$ \\ 
$30$ & $30$ & $100$ & $100$ & $0.101$ & $0.163$ & $0.146$ & $0.134$ & $0.087$ \\ 
$15$ & $30$ & $200$ & $100$ & $0.061$ & $0.156$ & $0.144$ & $0.088$ & $0.059$ \\ 
$30$ & $30$ & $200$ & $100$ & $0.073$ & $0.159$ & $0.146$ & $0.102$ & $0.063$ \\ 
$15$ & $15$ & $50$ & $200$ & $0.124$ & $0.114$ & $0.105$ & $0.194$ & $0.111$ \\ 
$30$ & $15$ & $50$ & $200$ & $0.144$ & $0.114$ & $0.105$ & $0.203$ & $0.114$ \\ 
$15$ & $15$ & $100$ & $200$ & $0.087$ & $0.112$ & $0.105$ & $0.152$ & $0.079$ \\ 
$30$ & $15$ & $100$ & $200$ & $0.102$ & $0.114$ & $0.107$ & $0.163$ & $0.085$ \\ 
$15$ & $15$ & $200$ & $200$ & $0.062$ & $0.110$ & $0.104$ & $0.114$ & $0.058$ \\ 
$30$ & $15$ & $200$ & $200$ & $0.073$ & $0.109$ & $0.104$ & $0.126$ & $0.061$ \\ 
$15$ & $30$ & $50$ & $200$ & $0.123$ & $0.113$ & $0.106$ & $0.194$ & $0.109$ \\ 
$30$ & $30$ & $50$ & $200$ & $0.142$ & $0.117$ & $0.107$ & $0.204$ & $0.113$ \\ 
$15$ & $30$ & $100$ & $200$ & $0.089$ & $0.116$ & $0.107$ & $0.151$ & $0.081$ \\ 
$30$ & $30$ & $100$ & $200$ & $0.103$ & $0.112$ & $0.104$ & $0.162$ & $0.086$ \\ 
$15$ & $30$ & $200$ & $200$ & $0.062$ & $0.112$ & $0.105$ & $0.113$ & $0.058$ \\ 
$30$ & $30$ & $200$ & $200$ & $0.073$ & $0.113$ & $0.104$ & $0.125$ & $0.062$ \\ 
\hline \\[-1.8ex] 
\end{tabular} } }
\end{table} 

\begin{table}[!htbp] \centering 
  \caption{Comparison of average empirical Kolmogorov--Smirnov distances computed over $1000$ simulation replications for gamma distributed failure times with constant hazard, varying censoring proportions and sample sizes for the RC and LBRC data (RC \%/SS, LBRC \%/SS, respectively). Estimates include the Kaplan--Meier estimator using only the RC data (KM (RC)), the adjusted Kaplan--Meier estimator using only the LBRC data while ignoring length-bias structure (KM (LTRC)), the survival function NPMLE using only the LBRC data (NPMLE (LBRC)), the survival function NPMLE using both RC and LBRC data (NPMLE RC+LBRC) and the misspecified density ratio model estimator (DRM) with $h(x) = \sqrt{x}$.} 
  \label{supp_misspec1_consthaz} 
  {   \renewcommand{\arraystretch}{1.3}   \resizebox{\textwidth}{!}{
\begin{tabular}{cccc|ccccc}
RC \% & LBRC \% & RC SS & LBRC SS & KM (RC) & KM (LTRC) & NPMLE (LBRC) & NPMLE (RC+LBRC) & DRM \\ \hline

$15$ & $15$ & $50$ & $50$ & $0.125$ & $0.204$ & $0.184$ & $0.149$ & $0.116$ \\ 
$30$ & $15$ & $50$ & $50$ & $0.139$ & $0.200$ & $0.183$ & $0.158$ & $0.124$ \\ 
$15$ & $15$ & $100$ & $50$ & $0.089$ & $0.195$ & $0.182$ & $0.107$ & $0.084$ \\ 
$30$ & $15$ & $100$ & $50$ & $0.097$ & $0.197$ & $0.181$ & $0.116$ & $0.089$ \\ 
$15$ & $15$ & $200$ & $50$ & $0.063$ & $0.196$ & $0.181$ & $0.075$ & $0.061$ \\ 
$30$ & $15$ & $200$ & $50$ & $0.070$ & $0.196$ & $0.180$ & $0.083$ & $0.066$ \\ 
$15$ & $30$ & $50$ & $50$ & $0.127$ & $0.187$ & $0.172$ & $0.148$ & $0.115$ \\ 
$30$ & $30$ & $50$ & $50$ & $0.139$ & $0.185$ & $0.172$ & $0.157$ & $0.122$ \\ 
$15$ & $30$ & $100$ & $50$ & $0.089$ & $0.193$ & $0.176$ & $0.107$ & $0.084$ \\ 
$30$ & $30$ & $100$ & $50$ & $0.099$ & $0.188$ & $0.174$ & $0.116$ & $0.090$ \\ 
$15$ & $30$ & $200$ & $50$ & $0.062$ & $0.190$ & $0.175$ & $0.074$ & $0.060$ \\ 
$30$ & $30$ & $200$ & $50$ & $0.070$ & $0.190$ & $0.173$ & $0.085$ & $0.065$ \\ 
$15$ & $15$ & $50$ & $100$ & $0.124$ & $0.144$ & $0.131$ & $0.171$ & $0.111$ \\ 
$30$ & $15$ & $50$ & $100$ & $0.138$ & $0.143$ & $0.131$ & $0.182$ & $0.117$ \\ 
$15$ & $15$ & $100$ & $100$ & $0.089$ & $0.142$ & $0.132$ & $0.129$ & $0.082$ \\ 
$30$ & $15$ & $100$ & $100$ & $0.098$ & $0.139$ & $0.128$ & $0.138$ & $0.086$ \\ 
$15$ & $15$ & $200$ & $100$ & $0.063$ & $0.140$ & $0.129$ & $0.093$ & $0.060$ \\ 
$30$ & $15$ & $200$ & $100$ & $0.068$ & $0.141$ & $0.130$ & $0.101$ & $0.062$ \\ 
$15$ & $30$ & $50$ & $100$ & $0.125$ & $0.140$ & $0.128$ & $0.171$ & $0.110$ \\ 
$30$ & $30$ & $50$ & $100$ & $0.138$ & $0.138$ & $0.127$ & $0.182$ & $0.114$ \\ 
$15$ & $30$ & $100$ & $100$ & $0.089$ & $0.136$ & $0.124$ & $0.131$ & $0.082$ \\ 
$30$ & $30$ & $100$ & $100$ & $0.098$ & $0.136$ & $0.127$ & $0.139$ & $0.085$ \\ 
$15$ & $30$ & $200$ & $100$ & $0.064$ & $0.139$ & $0.129$ & $0.095$ & $0.060$ \\ 
$30$ & $30$ & $200$ & $100$ & $0.069$ & $0.136$ & $0.126$ & $0.102$ & $0.062$ \\ 
$15$ & $15$ & $50$ & $200$ & $0.125$ & $0.102$ & $0.096$ & $0.208$ & $0.106$ \\ 
$30$ & $15$ & $50$ & $200$ & $0.138$ & $0.100$ & $0.094$ & $0.212$ & $0.110$ \\ 
$15$ & $15$ & $100$ & $200$ & $0.089$ & $0.102$ & $0.094$ & $0.161$ & $0.079$ \\ 
$30$ & $15$ & $100$ & $200$ & $0.098$ & $0.100$ & $0.094$ & $0.169$ & $0.083$ \\ 
$15$ & $15$ & $200$ & $200$ & $0.063$ & $0.100$ & $0.093$ & $0.119$ & $0.058$ \\ 
$30$ & $15$ & $200$ & $200$ & $0.069$ & $0.100$ & $0.094$ & $0.128$ & $0.061$ \\ 
$15$ & $30$ & $50$ & $200$ & $0.128$ & $0.097$ & $0.091$ & $0.206$ & $0.108$ \\ 
$30$ & $30$ & $50$ & $200$ & $0.138$ & $0.097$ & $0.090$ & $0.212$ & $0.110$ \\ 
$15$ & $30$ & $100$ & $200$ & $0.088$ & $0.095$ & $0.091$ & $0.161$ & $0.078$ \\ 
$30$ & $30$ & $100$ & $200$ & $0.098$ & $0.097$ & $0.090$ & $0.171$ & $0.083$ \\ 
$15$ & $30$ & $200$ & $200$ & $0.062$ & $0.097$ & $0.090$ & $0.120$ & $0.057$ \\ 
$30$ & $30$ & $200$ & $200$ & $0.069$ & $0.096$ & $0.090$ & $0.130$ & $0.060$ \\ 
\hline \\[-1.8ex] 
\end{tabular} } }
\end{table}

\begin{table}[!htbp] \centering 
  \caption{Comparison of average empirical Kolmogorov--Smirnov distances computed over $1000$ simulation replications for gamma distributed failure times with increasing hazard, varying censoring proportions and sample sizes for the RC and LBRC data (RC \%/SS, LBRC \%/SS, respectively). Estimates include the Kaplan--Meier estimator using only the RC data (KM (RC)), the adjusted Kaplan--Meier estimator using only the LBRC data while ignoring length-bias structure (KM (LTRC)), the survival function NPMLE using only the LBRC data (NPMLE (LBRC)), the survival function NPMLE using both RC and LBRC data (NPMLE RC+LBRC) and the misspecified density ratio model estimator (DRM) with $h(x) = \sqrt{x}$.} 
  \label{supp_misspec1_inchaz} 
  {   \renewcommand{\arraystretch}{1.3}   \resizebox{\textwidth}{!}{
\begin{tabular}{cccc|ccccc}
RC \% & LBRC \% & RC SS & LBRC SS & KM (RC) & KM (LTRC) & NPMLE (LBRC) & NPMLE (RC+LBRC) & DRM \\ \hline
$15$ & $15$ & $50$ & $50$ & $0.125$ & $0.169$ & $0.157$ & $0.143$ & $0.111$ \\ 
$30$ & $15$ & $50$ & $50$ & $0.141$ & $0.166$ & $0.156$ & $0.152$ & $0.120$ \\ 
$15$ & $15$ & $100$ & $50$ & $0.091$ & $0.165$ & $0.155$ & $0.105$ & $0.084$ \\ 
$30$ & $15$ & $100$ & $50$ & $0.098$ & $0.168$ & $0.157$ & $0.113$ & $0.088$ \\ 
$15$ & $15$ & $200$ & $50$ & $0.065$ & $0.168$ & $0.154$ & $0.075$ & $0.062$ \\ 
$30$ & $15$ & $200$ & $50$ & $0.069$ & $0.168$ & $0.155$ & $0.080$ & $0.064$ \\ 
$15$ & $30$ & $50$ & $50$ & $0.127$ & $0.172$ & $0.162$ & $0.143$ & $0.113$ \\ 
$30$ & $30$ & $50$ & $50$ & $0.136$ & $0.175$ & $0.161$ & $0.152$ & $0.118$ \\ 
$15$ & $30$ & $100$ & $50$ & $0.089$ & $0.173$ & $0.159$ & $0.104$ & $0.083$ \\ 
$30$ & $30$ & $100$ & $50$ & $0.098$ & $0.175$ & $0.161$ & $0.113$ & $0.090$ \\ 
$15$ & $30$ & $200$ & $50$ & $0.064$ & $0.172$ & $0.159$ & $0.073$ & $0.062$ \\ 
$30$ & $30$ & $200$ & $50$ & $0.070$ & $0.176$ & $0.162$ & $0.080$ & $0.066$ \\ 
$15$ & $15$ & $50$ & $100$ & $0.126$ & $0.118$ & $0.112$ & $0.167$ & $0.104$ \\ 
$30$ & $15$ & $50$ & $100$ & $0.137$ & $0.120$ & $0.111$ & $0.176$ & $0.111$ \\ 
$15$ & $15$ & $100$ & $100$ & $0.091$ & $0.118$ & $0.110$ & $0.125$ & $0.080$ \\ 
$30$ & $15$ & $100$ & $100$ & $0.098$ & $0.118$ & $0.111$ & $0.134$ & $0.084$ \\ 
$15$ & $15$ & $200$ & $100$ & $0.064$ & $0.119$ & $0.111$ & $0.090$ & $0.059$ \\ 
$30$ & $15$ & $200$ & $100$ & $0.069$ & $0.119$ & $0.111$ & $0.099$ & $0.062$ \\ 
$15$ & $30$ & $50$ & $100$ & $0.127$ & $0.122$ & $0.115$ & $0.166$ & $0.107$ \\ 
$30$ & $30$ & $50$ & $100$ & $0.135$ & $0.124$ & $0.115$ & $0.170$ & $0.110$ \\ 
$15$ & $30$ & $100$ & $100$ & $0.090$ & $0.126$ & $0.117$ & $0.126$ & $0.080$ \\ 
$30$ & $30$ & $100$ & $100$ & $0.097$ & $0.127$ & $0.118$ & $0.133$ & $0.084$ \\ 
$15$ & $30$ & $200$ & $100$ & $0.064$ & $0.120$ & $0.113$ & $0.089$ & $0.060$ \\ 
$30$ & $30$ & $200$ & $100$ & $0.070$ & $0.124$ & $0.116$ & $0.098$ & $0.064$ \\ 
$15$ & $15$ & $50$ & $200$ & $0.127$ & $0.084$ & $0.079$ & $0.192$ & $0.099$ \\ 
$30$ & $15$ & $50$ & $200$ & $0.138$ & $0.084$ & $0.079$ & $0.199$ & $0.101$ \\ 
$15$ & $15$ & $100$ & $200$ & $0.090$ & $0.083$ & $0.078$ & $0.155$ & $0.075$ \\ 
$30$ & $15$ & $100$ & $200$ & $0.099$ & $0.085$ & $0.080$ & $0.161$ & $0.079$ \\ 
$15$ & $15$ & $200$ & $200$ & $0.064$ & $0.085$ & $0.080$ & $0.117$ & $0.057$ \\ 
$30$ & $15$ & $200$ & $200$ & $0.070$ & $0.085$ & $0.080$ & $0.123$ & $0.060$ \\ 
$15$ & $30$ & $50$ & $200$ & $0.124$ & $0.088$ & $0.083$ & $0.191$ & $0.097$ \\ 
$30$ & $30$ & $50$ & $200$ & $0.138$ & $0.088$ & $0.083$ & $0.196$ & $0.103$ \\ 
$15$ & $30$ & $100$ & $200$ & $0.088$ & $0.087$ & $0.081$ & $0.152$ & $0.074$ \\ 
$30$ & $30$ & $100$ & $200$ & $0.097$ & $0.086$ & $0.081$ & $0.161$ & $0.079$ \\ 
$15$ & $30$ & $200$ & $200$ & $0.065$ & $0.088$ & $0.083$ & $0.114$ & $0.058$ \\ 
$30$ & $30$ & $200$ & $200$ & $0.070$ & $0.087$ & $0.082$ & $0.122$ & $0.060$ \\ 
\hline \\[-1.8ex] 
\end{tabular} } }
\end{table}

\begin{table}[!htbp] \centering 
  \caption{Comparison of average empirical Kolmogorov--Smirnov distances computed over $1000$ simulation replications for gamma distributed failure times with decreasing hazard, varying censoring proportions and sample sizes for the RC and LBRC data (RC \%/SS, LBRC \%/SS, respectively). Estimates include the Kaplan--Meier estimator using only the RC data (KM (RC)), the adjusted Kaplan--Meier estimator using only the LBRC data while ignoring length-bias structure (KM (LTRC)), the survival function NPMLE using only the LBRC data (NPMLE (LBRC)), the survival function NPMLE using both RC and LBRC data (NPMLE RC+LBRC) and the incorrectly specified density ratio model estimator (DRM) with $h(x) = x$.} 
  \label{supp_misspec2_dechaz} 
  {   \renewcommand{\arraystretch}{1.3}   \resizebox{\textwidth}{!}{
\begin{tabular}{cccc|ccccc}
RC \% & LBRC \% & RC SS & LBRC SS & KM (RC) & KM (LTRC) & NPMLE (LBRC) & NPMLE (RC+LBRC) & DRM \\ \hline

$15$ & $15$ & $50$ & $50$ & $0.124$ & $0.209$ & $0.193$ & $0.140$ & $0.116$ \\ 
$30$ & $15$ & $50$ & $50$ & $0.145$ & $0.214$ & $0.191$ & $0.150$ & $0.125$ \\ 
$15$ & $15$ & $100$ & $50$ & $0.088$ & $0.211$ & $0.193$ & $0.103$ & $0.084$ \\ 
$30$ & $15$ & $100$ & $50$ & $0.102$ & $0.211$ & $0.192$ & $0.116$ & $0.091$ \\ 
$15$ & $15$ & $200$ & $50$ & $0.062$ & $0.212$ & $0.192$ & $0.074$ & $0.060$ \\ 
$30$ & $15$ & $200$ & $50$ & $0.074$ & $0.212$ & $0.192$ & $0.085$ & $0.066$ \\ 
$15$ & $30$ & $50$ & $50$ & $0.125$ & $0.219$ & $0.199$ & $0.142$ & $0.118$ \\ 
$30$ & $30$ & $50$ & $50$ & $0.145$ & $0.218$ & $0.197$ & $0.153$ & $0.126$ \\ 
$15$ & $30$ & $100$ & $50$ & $0.088$ & $0.218$ & $0.198$ & $0.106$ & $0.085$ \\ 
$30$ & $30$ & $100$ & $50$ & $0.100$ & $0.218$ & $0.196$ & $0.113$ & $0.089$ \\ 
$15$ & $30$ & $200$ & $50$ & $0.062$ & $0.221$ & $0.197$ & $0.073$ & $0.060$ \\ 
$30$ & $30$ & $200$ & $50$ & $0.074$ & $0.218$ & $0.195$ & $0.085$ & $0.066$ \\ 
$15$ & $15$ & $50$ & $100$ & $0.122$ & $0.155$ & $0.143$ & $0.164$ & $0.114$ \\ 
$30$ & $15$ & $50$ & $100$ & $0.143$ & $0.155$ & $0.141$ & $0.173$ & $0.120$ \\ 
$15$ & $15$ & $100$ & $100$ & $0.087$ & $0.157$ & $0.142$ & $0.124$ & $0.083$ \\ 
$30$ & $15$ & $100$ & $100$ & $0.103$ & $0.154$ & $0.141$ & $0.135$ & $0.088$ \\ 
$15$ & $15$ & $200$ & $100$ & $0.062$ & $0.156$ & $0.141$ & $0.090$ & $0.060$ \\ 
$30$ & $15$ & $200$ & $100$ & $0.073$ & $0.154$ & $0.142$ & $0.102$ & $0.064$ \\ 
$15$ & $30$ & $50$ & $100$ & $0.125$ & $0.160$ & $0.147$ & $0.164$ & $0.117$ \\ 
$30$ & $30$ & $50$ & $100$ & $0.144$ & $0.161$ & $0.146$ & $0.172$ & $0.122$ \\ 
$15$ & $30$ & $100$ & $100$ & $0.088$ & $0.159$ & $0.146$ & $0.124$ & $0.084$ \\ 
$30$ & $30$ & $100$ & $100$ & $0.101$ & $0.162$ & $0.146$ & $0.134$ & $0.088$ \\ 
$15$ & $30$ & $200$ & $100$ & $0.061$ & $0.157$ & $0.144$ & $0.088$ & $0.059$ \\ 
$30$ & $30$ & $200$ & $100$ & $0.073$ & $0.159$ & $0.145$ & $0.101$ & $0.064$ \\ 
$15$ & $15$ & $50$ & $200$ & $0.124$ & $0.114$ & $0.105$ & $0.193$ & $0.113$ \\ 
$30$ & $15$ & $50$ & $200$ & $0.144$ & $0.114$ & $0.105$ & $0.203$ & $0.117$ \\ 
$15$ & $15$ & $100$ & $200$ & $0.087$ & $0.113$ & $0.105$ & $0.151$ & $0.080$ \\ 
$30$ & $15$ & $100$ & $200$ & $0.102$ & $0.115$ & $0.107$ & $0.163$ & $0.087$ \\ 
$15$ & $15$ & $200$ & $200$ & $0.062$ & $0.110$ & $0.104$ & $0.114$ & $0.059$ \\ 
$30$ & $15$ & $200$ & $200$ & $0.073$ & $0.109$ & $0.104$ & $0.126$ & $0.063$ \\ 
$15$ & $30$ & $50$ & $200$ & $0.123$ & $0.113$ & $0.106$ & $0.194$ & $0.112$ \\ 
$30$ & $30$ & $50$ & $200$ & $0.142$ & $0.117$ & $0.107$ & $0.204$ & $0.115$ \\ 
$15$ & $30$ & $100$ & $200$ & $0.089$ & $0.117$ & $0.107$ & $0.151$ & $0.083$ \\ 
$30$ & $30$ & $100$ & $200$ & $0.102$ & $0.113$ & $0.104$ & $0.162$ & $0.088$ \\ 
$15$ & $30$ & $200$ & $200$ & $0.062$ & $0.112$ & $0.105$ & $0.113$ & $0.060$ \\ 
$30$ & $30$ & $200$ & $200$ & $0.073$ & $0.113$ & $0.104$ & $0.125$ & $0.063$ \\ 
\hline \\[-1.8ex] 
\end{tabular} } }
\end{table}

\begin{table}[!htbp] \centering 
  \caption{Comparison of average empirical Kolmogorov--Smirnov distances computed over $1000$ simulation replications for gamma distributed failure times with constant hazard, varying censoring proportions and sample sizes for the RC and LBRC data (RC \%/SS, LBRC \%/SS, respectively). Estimates include the Kaplan--Meier estimator using only the RC data (KM (RC)), the adjusted Kaplan--Meier estimator using only the LBRC data while ignoring length-bias structure (KM (LTRC)), the survival function NPMLE using only the LBRC data (NPMLE (LBRC)), the survival function NPMLE using both RC and LBRC data (NPMLE RC+LBRC) and the incorrectly specified density ratio model estimator (DRM) with $h(x) = x$. } 
  \label{supp_misspec2_consthaz} 
  {   \renewcommand{\arraystretch}{1.3}   \resizebox{\textwidth}{!}{
\begin{tabular}{cccc|ccccc}
RC \% & LBRC \% & RC SS & LBRC SS & KM (RC) & KM (LTRC) & NPMLE (LBRC) & NPMLE (RC+LBRC) & DRM \\ \hline

$15$ & $15$ & $50$ & $50$ & $0.125$ & $0.205$ & $0.185$ & $0.149$ & $0.116$ \\ 
$30$ & $15$ & $50$ & $50$ & $0.139$ & $0.200$ & $0.183$ & $0.158$ & $0.125$ \\ 
$15$ & $15$ & $100$ & $50$ & $0.089$ & $0.196$ & $0.182$ & $0.107$ & $0.084$ \\ 
$30$ & $15$ & $100$ & $50$ & $0.097$ & $0.197$ & $0.181$ & $0.116$ & $0.089$ \\ 
$15$ & $15$ & $200$ & $50$ & $0.063$ & $0.196$ & $0.180$ & $0.074$ & $0.061$ \\ 
$30$ & $15$ & $200$ & $50$ & $0.070$ & $0.195$ & $0.179$ & $0.083$ & $0.066$ \\ 
$15$ & $30$ & $50$ & $50$ & $0.127$ & $0.187$ & $0.171$ & $0.148$ & $0.115$ \\ 
$30$ & $30$ & $50$ & $50$ & $0.139$ & $0.184$ & $0.172$ & $0.158$ & $0.124$ \\ 
$15$ & $30$ & $100$ & $50$ & $0.089$ & $0.194$ & $0.177$ & $0.107$ & $0.085$ \\ 
$30$ & $30$ & $100$ & $50$ & $0.099$ & $0.187$ & $0.173$ & $0.116$ & $0.090$ \\ 
$15$ & $30$ & $200$ & $50$ & $0.062$ & $0.190$ & $0.174$ & $0.074$ & $0.060$ \\ 
$30$ & $30$ & $200$ & $50$ & $0.069$ & $0.190$ & $0.173$ & $0.085$ & $0.065$ \\ 
$15$ & $15$ & $50$ & $100$ & $0.124$ & $0.143$ & $0.130$ & $0.171$ & $0.112$ \\ 
$30$ & $15$ & $50$ & $100$ & $0.138$ & $0.142$ & $0.130$ & $0.182$ & $0.118$ \\ 
$15$ & $15$ & $100$ & $100$ & $0.090$ & $0.142$ & $0.132$ & $0.129$ & $0.083$ \\ 
$30$ & $15$ & $100$ & $100$ & $0.097$ & $0.139$ & $0.128$ & $0.138$ & $0.087$ \\ 
$15$ & $15$ & $200$ & $100$ & $0.063$ & $0.139$ & $0.129$ & $0.093$ & $0.061$ \\ 
$30$ & $15$ & $200$ & $100$ & $0.068$ & $0.142$ & $0.130$ & $0.101$ & $0.063$ \\ 
$15$ & $30$ & $50$ & $100$ & $0.125$ & $0.139$ & $0.128$ & $0.171$ & $0.110$ \\ 
$30$ & $30$ & $50$ & $100$ & $0.138$ & $0.137$ & $0.127$ & $0.183$ & $0.117$ \\ 
$15$ & $30$ & $100$ & $100$ & $0.089$ & $0.135$ & $0.124$ & $0.130$ & $0.083$ \\ 
$30$ & $30$ & $100$ & $100$ & $0.098$ & $0.135$ & $0.127$ & $0.139$ & $0.086$ \\ 
$15$ & $30$ & $200$ & $100$ & $0.064$ & $0.140$ & $0.129$ & $0.094$ & $0.061$ \\ 
$30$ & $30$ & $200$ & $100$ & $0.068$ & $0.136$ & $0.125$ & $0.103$ & $0.063$ \\ 
$15$ & $15$ & $50$ & $200$ & $0.125$ & $0.101$ & $0.096$ & $0.208$ & $0.111$ \\ 
$30$ & $15$ & $50$ & $200$ & $0.138$ & $0.100$ & $0.094$ & $0.212$ & $0.114$ \\ 
$15$ & $15$ & $100$ & $200$ & $0.089$ & $0.102$ & $0.094$ & $0.160$ & $0.082$ \\ 
$30$ & $15$ & $100$ & $200$ & $0.098$ & $0.101$ & $0.094$ & $0.169$ & $0.086$ \\ 
$15$ & $15$ & $200$ & $200$ & $0.062$ & $0.099$ & $0.092$ & $0.119$ & $0.059$ \\ 
$30$ & $15$ & $200$ & $200$ & $0.069$ & $0.100$ & $0.094$ & $0.128$ & $0.063$ \\ 
$15$ & $30$ & $50$ & $200$ & $0.127$ & $0.098$ & $0.092$ & $0.206$ & $0.111$ \\ 
$30$ & $30$ & $50$ & $200$ & $0.138$ & $0.097$ & $0.090$ & $0.213$ & $0.112$ \\ 
$15$ & $30$ & $100$ & $200$ & $0.088$ & $0.095$ & $0.091$ & $0.161$ & $0.080$ \\ 
$30$ & $30$ & $100$ & $200$ & $0.098$ & $0.098$ & $0.090$ & $0.171$ & $0.086$ \\ 
$15$ & $30$ & $200$ & $200$ & $0.062$ & $0.097$ & $0.090$ & $0.120$ & $0.059$ \\ 
$30$ & $30$ & $200$ & $200$ & $0.069$ & $0.097$ & $0.091$ & $0.130$ & $0.062$ \\ 
\hline \\[-1.8ex] 
\end{tabular} } }
\end{table}

\begin{table}[!htbp] \centering 
  \caption{Comparison of average empirical Kolmogorov--Smirnov distances computed over $1000$ simulation replications for gamma distributed failure times with increasing hazard, varying censoring proportions and sample sizes for the RC and LBRC data (RC \%/SS, LBRC \%/SS, respectively). Estimates include the Kaplan--Meier estimator using only the RC data (KM (RC)), the adjusted Kaplan--Meier estimator using only the LBRC data while ignoring length-bias structure (KM (LTRC)), the survival function NPMLE using only the LBRC data (NPMLE (LBRC)), the survival function NPMLE using both RC and LBRC data (NPMLE RC+LBRC) and the incorrectly specified density ratio model estimator (DRM) with $h(x) = x$.} 
  \label{supp_misspec2_inchaz} 
  {   \renewcommand{\arraystretch}{1.3}   \resizebox{\textwidth}{!}{
\begin{tabular}{cccc|ccccc}
RC \% & LBRC \% & RC SS & LBRC SS & KM (RC) & KM (LTRC) & NPMLE (LBRC) & NPMLE (RC+LBRC) & DRM \\ \hline 
$15$ & $15$ & $50$ & $50$ & $0.126$ & $0.169$ & $0.157$ & $0.144$ & $0.112$ \\ 
$30$ & $15$ & $50$ & $50$ & $0.141$ & $0.167$ & $0.156$ & $0.152$ & $0.121$ \\ 
$15$ & $15$ & $100$ & $50$ & $0.091$ & $0.164$ & $0.154$ & $0.105$ & $0.084$ \\ 
$30$ & $15$ & $100$ & $50$ & $0.098$ & $0.168$ & $0.158$ & $0.113$ & $0.088$ \\ 
$15$ & $15$ & $200$ & $50$ & $0.065$ & $0.167$ & $0.155$ & $0.075$ & $0.062$ \\ 
$30$ & $15$ & $200$ & $50$ & $0.069$ & $0.168$ & $0.155$ & $0.080$ & $0.065$ \\ 
$15$ & $30$ & $50$ & $50$ & $0.128$ & $0.172$ & $0.162$ & $0.142$ & $0.113$ \\ 
$30$ & $30$ & $50$ & $50$ & $0.137$ & $0.175$ & $0.160$ & $0.152$ & $0.119$ \\ 
$15$ & $30$ & $100$ & $50$ & $0.088$ & $0.173$ & $0.160$ & $0.104$ & $0.083$ \\ 
$30$ & $30$ & $100$ & $50$ & $0.098$ & $0.174$ & $0.161$ & $0.113$ & $0.090$ \\ 
$15$ & $30$ & $200$ & $50$ & $0.064$ & $0.172$ & $0.159$ & $0.073$ & $0.062$ \\ 
$30$ & $30$ & $200$ & $50$ & $0.070$ & $0.175$ & $0.162$ & $0.080$ & $0.066$ \\ 
$15$ & $15$ & $50$ & $100$ & $0.126$ & $0.117$ & $0.112$ & $0.167$ & $0.106$ \\ 
$30$ & $15$ & $50$ & $100$ & $0.136$ & $0.121$ & $0.112$ & $0.176$ & $0.113$ \\ 
$15$ & $15$ & $100$ & $100$ & $0.091$ & $0.118$ & $0.111$ & $0.126$ & $0.082$ \\ 
$30$ & $15$ & $100$ & $100$ & $0.098$ & $0.118$ & $0.111$ & $0.134$ & $0.085$ \\ 
$15$ & $15$ & $200$ & $100$ & $0.064$ & $0.118$ & $0.111$ & $0.090$ & $0.060$ \\ 
$30$ & $15$ & $200$ & $100$ & $0.068$ & $0.119$ & $0.111$ & $0.099$ & $0.063$ \\ 
$15$ & $30$ & $50$ & $100$ & $0.127$ & $0.122$ & $0.115$ & $0.166$ & $0.108$ \\ 
$30$ & $30$ & $50$ & $100$ & $0.135$ & $0.124$ & $0.115$ & $0.170$ & $0.112$ \\ 
$15$ & $30$ & $100$ & $100$ & $0.090$ & $0.126$ & $0.117$ & $0.126$ & $0.081$ \\ 
$30$ & $30$ & $100$ & $100$ & $0.097$ & $0.127$ & $0.118$ & $0.133$ & $0.086$ \\ 
$15$ & $30$ & $200$ & $100$ & $0.064$ & $0.120$ & $0.113$ & $0.088$ & $0.060$ \\ 
$30$ & $30$ & $200$ & $100$ & $0.070$ & $0.124$ & $0.116$ & $0.098$ & $0.064$ \\ 
$15$ & $15$ & $50$ & $200$ & $0.127$ & $0.084$ & $0.079$ & $0.192$ & $0.102$ \\ 
$30$ & $15$ & $50$ & $200$ & $0.137$ & $0.085$ & $0.080$ & $0.199$ & $0.103$ \\ 
$15$ & $15$ & $100$ & $200$ & $0.090$ & $0.083$ & $0.078$ & $0.155$ & $0.078$ \\ 
$30$ & $15$ & $100$ & $200$ & $0.098$ & $0.085$ & $0.080$ & $0.161$ & $0.080$ \\ 
$15$ & $15$ & $200$ & $200$ & $0.064$ & $0.085$ & $0.080$ & $0.117$ & $0.059$ \\ 
$30$ & $15$ & $200$ & $200$ & $0.070$ & $0.085$ & $0.081$ & $0.124$ & $0.062$ \\ 
$15$ & $30$ & $50$ & $200$ & $0.124$ & $0.088$ & $0.083$ & $0.192$ & $0.100$ \\ 
$30$ & $30$ & $50$ & $200$ & $0.138$ & $0.089$ & $0.083$ & $0.196$ & $0.106$ \\ 
$15$ & $30$ & $100$ & $200$ & $0.089$ & $0.087$ & $0.082$ & $0.152$ & $0.077$ \\ 
$30$ & $30$ & $100$ & $200$ & $0.097$ & $0.086$ & $0.081$ & $0.161$ & $0.082$ \\ 
$15$ & $30$ & $200$ & $200$ & $0.064$ & $0.088$ & $0.083$ & $0.115$ & $0.059$ \\ 
$30$ & $30$ & $200$ & $200$ & $0.070$ & $0.087$ & $0.081$ & $0.122$ & $0.061$ \\ 
\hline \\[-1.8ex] 
\end{tabular} } }
\end{table}

\begin{table}[!htbp] \centering 
\caption{Comparison of average empirical Kolmogorov--Smirnov distances computed over $1000$ simulation replications for gamma distributed failure times with decreasing hazard, varying censoring proportions and sample sizes for the RC and LBRC data (RC \%/SS, LBRC \%/SS, respectively). The RC and LBRC failure times are assumed to be drawn from the same distribution. Estimates include the Kaplan--Meier estimator using only the RC data (KM (RC)), the adjusted Kaplan--Meier estimator using only the LBRC data while ignoring length-bias structure (KM (LTRC)), the survival function NPMLE using only the LBRC data (NPMLE (LBRC)), the survival function NPMLE using both RC and LBRC data (NPMLE RC+LBRC) and the density ratio model estimator (DRM) with $h(x) = \log(x)$.} 
  \label{supp_identical_dechaz} 
  {   \renewcommand{\arraystretch}{1.3}   \resizebox{\textwidth}{!}{
\begin{tabular}{cccc|ccccc}
RC \% & LBRC \% & RC SS & LBRC SS & KM (RC) & KM (LTRC) & NPMLE (LBRC) & NPMLE (RC+LBRC) & DRM \\ \hline 
$15$ & $15$ & $500$ & $500$ & $0.039$ & $0.182$ & $0.150$ & $0.032$ & $0.034$ \\ 
$30$ & $15$ & $500$ & $500$ & $0.046$ & $0.179$ & $0.152$ & $0.033$ & $0.035$ \\ 
$15$ & $15$ & $1,000$ & $500$ & $0.028$ & $0.183$ & $0.153$ & $0.025$ & $0.025$ \\ 
$30$ & $15$ & $1,000$ & $500$ & $0.033$ & $0.185$ & $0.153$ & $0.025$ & $0.026$ \\ 
$15$ & $15$ & $2,000$ & $500$ & $0.020$ & $0.186$ & $0.154$ & $0.018$ & $0.018$ \\ 
$30$ & $15$ & $2,000$ & $500$ & $0.024$ & $0.182$ & $0.154$ & $0.019$ & $0.020$ \\ 
$15$ & $30$ & $500$ & $500$ & $0.040$ & $0.187$ & $0.152$ & $0.033$ & $0.035$ \\ 
$30$ & $30$ & $500$ & $500$ & $0.047$ & $0.180$ & $0.151$ & $0.033$ & $0.035$ \\ 
$15$ & $30$ & $1,000$ & $500$ & $0.028$ & $0.188$ & $0.152$ & $0.026$ & $0.026$ \\ 
$30$ & $30$ & $1,000$ & $500$ & $0.033$ & $0.182$ & $0.152$ & $0.025$ & $0.026$ \\ 
$15$ & $30$ & $2,000$ & $500$ & $0.020$ & $0.182$ & $0.156$ & $0.019$ & $0.019$ \\ 
$30$ & $30$ & $2,000$ & $500$ & $0.024$ & $0.177$ & $0.150$ & $0.019$ & $0.020$ \\ 
$15$ & $15$ & $500$ & $1,000$ & $0.040$ & $0.144$ & $0.117$ & $0.031$ & $0.032$ \\ 
$30$ & $15$ & $500$ & $1,000$ & $0.046$ & $0.148$ & $0.118$ & $0.031$ & $0.033$ \\ 
$15$ & $15$ & $1,000$ & $1,000$ & $0.027$ & $0.139$ & $0.116$ & $0.023$ & $0.024$ \\ 
$30$ & $15$ & $1,000$ & $1,000$ & $0.033$ & $0.141$ & $0.114$ & $0.023$ & $0.024$ \\ 
$15$ & $15$ & $2,000$ & $1,000$ & $0.020$ & $0.142$ & $0.113$ & $0.017$ & $0.018$ \\ 
$30$ & $15$ & $2,000$ & $1,000$ & $0.024$ & $0.139$ & $0.116$ & $0.018$ & $0.019$ \\ 
$15$ & $30$ & $500$ & $1,000$ & $0.039$ & $0.146$ & $0.115$ & $0.030$ & $0.031$ \\ 
$30$ & $30$ & $500$ & $1,000$ & $0.046$ & $0.145$ & $0.117$ & $0.030$ & $0.033$ \\ 
$15$ & $30$ & $1,000$ & $1,000$ & $0.028$ & $0.146$ & $0.116$ & $0.025$ & $0.025$ \\ 
$30$ & $30$ & $1,000$ & $1,000$ & $0.033$ & $0.143$ & $0.116$ & $0.023$ & $0.024$ \\ 
$15$ & $30$ & $2,000$ & $1,000$ & $0.020$ & $0.148$ & $0.114$ & $0.018$ & $0.018$ \\ 
$30$ & $30$ & $2,000$ & $1,000$ & $0.023$ & $0.145$ & $0.117$ & $0.018$ & $0.018$ \\ 
$15$ & $15$ & $500$ & $2,000$ & $0.039$ & $0.122$ & $0.092$ & $0.027$ & $0.030$ \\ 
$30$ & $15$ & $500$ & $2,000$ & $0.046$ & $0.118$ & $0.092$ & $0.027$ & $0.030$ \\ 
$15$ & $15$ & $1,000$ & $2,000$ & $0.028$ & $0.113$ & $0.093$ & $0.021$ & $0.023$ \\ 
$30$ & $15$ & $1,000$ & $2,000$ & $0.033$ & $0.110$ & $0.090$ & $0.021$ & $0.023$ \\ 
$15$ & $15$ & $2,000$ & $2,000$ & $0.020$ & $0.118$ & $0.093$ & $0.017$ & $0.017$ \\ 
$30$ & $15$ & $2,000$ & $2,000$ & $0.023$ & $0.114$ & $0.092$ & $0.016$ & $0.017$ \\ 
$15$ & $30$ & $500$ & $2,000$ & $0.039$ & $0.112$ & $0.094$ & $0.026$ & $0.029$ \\ 
$30$ & $30$ & $500$ & $2,000$ & $0.046$ & $0.115$ & $0.092$ & $0.027$ & $0.030$ \\ 
$15$ & $30$ & $1,000$ & $2,000$ & $0.028$ & $0.116$ & $0.093$ & $0.021$ & $0.023$ \\ 
$30$ & $30$ & $1,000$ & $2,000$ & $0.033$ & $0.114$ & $0.092$ & $0.021$ & $0.024$ \\ 
$15$ & $30$ & $2,000$ & $2,000$ & $0.020$ & $0.112$ & $0.092$ & $0.018$ & $0.018$ \\ 
$30$ & $30$ & $2,000$ & $2,000$ & $0.023$ & $0.110$ & $0.092$ & $0.016$ & $0.018$ \\ 
\hline 
\end{tabular} } }
\end{table}

\begin{table}[!htbp] \centering 
\caption{Comparison of average empirical Kolmogorov--Smirnov distances computed over $1000$ simulation replications for gamma distributed failure times with constant hazard, varying censoring proportions and sample sizes for the RC and LBRC data (RC \%/SS, LBRC \%/SS, respectively). The RC and LBRC failure times are assumed to be drawn from the same distribution. Estimates include the Kaplan--Meier estimator using only the RC data (KM (RC)), the adjusted Kaplan--Meier estimator using only the LBRC data while ignoring length-bias structure (KM (LTRC)), the survival function NPMLE using only the LBRC data (NPMLE (LBRC)), the survival function NPMLE using both RC and LBRC data (NPMLE RC+LBRC) and the density ratio model estimator (DRM) with $h(x) = \log(x)$.}  
  \label{supp_identical_consthaz} 
  {   \renewcommand{\arraystretch}{1.3}   \resizebox{\textwidth}{!}{
\begin{tabular}{cccc|ccccc}
RC \% & LBRC \% & RC SS & LBRC SS & KM (RC) & KM (LTRC) & NPMLE (LBRC) & NPMLE (RC+LBRC) & DRM \\ \hline 
$15$ & $15$ & $500$ & $500$ & $0.040$ & $0.098$ & $0.091$ & $0.032$ & $0.034$ \\ 
$30$ & $15$ & $500$ & $500$ & $0.044$ & $0.098$ & $0.091$ & $0.032$ & $0.035$ \\ 
$15$ & $15$ & $1,000$ & $500$ & $0.028$ & $0.097$ & $0.089$ & $0.025$ & $0.025$ \\ 
$30$ & $15$ & $1,000$ & $500$ & $0.031$ & $0.097$ & $0.089$ & $0.026$ & $0.027$ \\ 
$15$ & $15$ & $2,000$ & $500$ & $0.020$ & $0.102$ & $0.092$ & $0.019$ & $0.019$ \\ 
$30$ & $15$ & $2,000$ & $500$ & $0.022$ & $0.100$ & $0.092$ & $0.019$ & $0.020$ \\ 
$15$ & $30$ & $500$ & $500$ & $0.040$ & $0.098$ & $0.091$ & $0.032$ & $0.034$ \\ 
$30$ & $30$ & $500$ & $500$ & $0.044$ & $0.099$ & $0.091$ & $0.033$ & $0.035$ \\ 
$15$ & $30$ & $1,000$ & $500$ & $0.028$ & $0.099$ & $0.092$ & $0.026$ & $0.025$ \\ 
$30$ & $30$ & $1,000$ & $500$ & $0.031$ & $0.100$ & $0.092$ & $0.026$ & $0.026$ \\ 
$15$ & $30$ & $2,000$ & $500$ & $0.020$ & $0.099$ & $0.091$ & $0.019$ & $0.019$ \\ 
$30$ & $30$ & $2,000$ & $500$ & $0.022$ & $0.100$ & $0.092$ & $0.019$ & $0.020$ \\ 
$15$ & $15$ & $500$ & $1,000$ & $0.040$ & $0.076$ & $0.067$ & $0.029$ & $0.031$ \\ 
$30$ & $15$ & $500$ & $1,000$ & $0.043$ & $0.070$ & $0.064$ & $0.029$ & $0.031$ \\ 
$15$ & $15$ & $1,000$ & $1,000$ & $0.028$ & $0.071$ & $0.066$ & $0.023$ & $0.024$ \\ 
$30$ & $15$ & $1,000$ & $1,000$ & $0.031$ & $0.072$ & $0.066$ & $0.023$ & $0.025$ \\ 
$15$ & $15$ & $2,000$ & $1,000$ & $0.020$ & $0.073$ & $0.066$ & $0.018$ & $0.018$ \\ 
$30$ & $15$ & $2,000$ & $1,000$ & $0.022$ & $0.073$ & $0.066$ & $0.018$ & $0.019$ \\ 
$15$ & $30$ & $500$ & $1,000$ & $0.040$ & $0.072$ & $0.066$ & $0.029$ & $0.031$ \\ 
$30$ & $30$ & $500$ & $1,000$ & $0.045$ & $0.072$ & $0.066$ & $0.029$ & $0.033$ \\ 
$15$ & $30$ & $1,000$ & $1,000$ & $0.028$ & $0.071$ & $0.066$ & $0.024$ & $0.024$ \\ 
$30$ & $30$ & $1,000$ & $1,000$ & $0.031$ & $0.075$ & $0.068$ & $0.023$ & $0.025$ \\ 
$15$ & $30$ & $2,000$ & $1,000$ & $0.020$ & $0.074$ & $0.067$ & $0.019$ & $0.018$ \\ 
$30$ & $30$ & $2,000$ & $1,000$ & $0.022$ & $0.072$ & $0.067$ & $0.018$ & $0.019$ \\ 
$15$ & $15$ & $500$ & $2,000$ & $0.039$ & $0.055$ & $0.048$ & $0.024$ & $0.028$ \\ 
$30$ & $15$ & $500$ & $2,000$ & $0.044$ & $0.053$ & $0.048$ & $0.024$ & $0.029$ \\ 
$15$ & $15$ & $1,000$ & $2,000$ & $0.028$ & $0.053$ & $0.047$ & $0.020$ & $0.021$ \\ 
$30$ & $15$ & $1,000$ & $2,000$ & $0.031$ & $0.053$ & $0.047$ & $0.020$ & $0.022$ \\ 
$15$ & $15$ & $2,000$ & $2,000$ & $0.020$ & $0.054$ & $0.048$ & $0.016$ & $0.017$ \\ 
$30$ & $15$ & $2,000$ & $2,000$ & $0.022$ & $0.052$ & $0.048$ & $0.016$ & $0.017$ \\ 
$15$ & $30$ & $500$ & $2,000$ & $0.040$ & $0.053$ & $0.048$ & $0.024$ & $0.028$ \\ 
$30$ & $30$ & $500$ & $2,000$ & $0.044$ & $0.054$ & $0.048$ & $0.024$ & $0.029$ \\ 
$15$ & $30$ & $1,000$ & $2,000$ & $0.028$ & $0.054$ & $0.049$ & $0.021$ & $0.022$ \\ 
$30$ & $30$ & $1,000$ & $2,000$ & $0.031$ & $0.053$ & $0.049$ & $0.020$ & $0.022$ \\ 
$15$ & $30$ & $2,000$ & $2,000$ & $0.020$ & $0.056$ & $0.048$ & $0.016$ & $0.017$ \\ 
$30$ & $30$ & $2,000$ & $2,000$ & $0.022$ & $0.053$ & $0.048$ & $0.016$ & $0.018$ \\ 
\hline \\[-1.8ex] 
\end{tabular} } }
\end{table}

\begin{table}[!htbp] \centering 
\caption{Comparison of average empirical Kolmogorov--Smirnov distances computed over $1000$ simulation replications for gamma distributed failure times with increasing hazard, varying censoring proportions and sample sizes for the RC and LBRC data (RC \%/SS, LBRC \%/SS, respectively). The RC and LBRC failure times are assumed to be drawn from the same distribution. Estimates include the Kaplan--Meier estimator using only the RC data (KM (RC)), the adjusted Kaplan--Meier estimator using only the LBRC data while ignoring length-bias structure (KM (LTRC)), the survival function NPMLE using only the LBRC data (NPMLE (LBRC)), the survival function NPMLE using both RC and LBRC data (NPMLE RC+LBRC) and the density ratio model estimator (DRM) with $h(x) = \log(x)$.}  
  \label{supp_identical_inchaz} 
  {   \renewcommand{\arraystretch}{1.3}   \resizebox{\textwidth}{!}{
\begin{tabular}{cccc|ccccc}
RC \% & LBRC \% & RC SS & LBRC SS & KM (RC) & KM (LTRC) & NPMLE (LBRC) & NPMLE (RC+LBRC) & DRM \\ \hline 
$15$ & $15$ & $500$ & $500$ & $0.041$ & $0.062$ & $0.059$ & $0.032$ & $0.033$ \\ 
$30$ & $15$ & $500$ & $500$ & $0.044$ & $0.064$ & $0.059$ & $0.032$ & $0.034$ \\ 
$15$ & $15$ & $1,000$ & $500$ & $0.029$ & $0.062$ & $0.059$ & $0.025$ & $0.025$ \\ 
$30$ & $15$ & $1,000$ & $500$ & $0.031$ & $0.062$ & $0.059$ & $0.025$ & $0.026$ \\ 
$15$ & $15$ & $2,000$ & $500$ & $0.020$ & $0.062$ & $0.059$ & $0.018$ & $0.018$ \\ 
$30$ & $15$ & $2,000$ & $500$ & $0.022$ & $0.062$ & $0.058$ & $0.020$ & $0.020$ \\ 
$15$ & $30$ & $500$ & $500$ & $0.041$ & $0.062$ & $0.059$ & $0.032$ & $0.034$ \\ 
$30$ & $30$ & $500$ & $500$ & $0.044$ & $0.063$ & $0.059$ & $0.032$ & $0.035$ \\ 
$15$ & $30$ & $1,000$ & $500$ & $0.028$ & $0.064$ & $0.061$ & $0.025$ & $0.025$ \\ 
$30$ & $30$ & $1,000$ & $500$ & $0.031$ & $0.064$ & $0.060$ & $0.026$ & $0.027$ \\ 
$15$ & $30$ & $2,000$ & $500$ & $0.020$ & $0.063$ & $0.059$ & $0.019$ & $0.019$ \\ 
$30$ & $30$ & $2,000$ & $500$ & $0.022$ & $0.063$ & $0.060$ & $0.020$ & $0.020$ \\ 
$15$ & $15$ & $500$ & $1,000$ & $0.041$ & $0.044$ & $0.042$ & $0.026$ & $0.030$ \\ 
$30$ & $15$ & $500$ & $1,000$ & $0.043$ & $0.044$ & $0.042$ & $0.026$ & $0.031$ \\ 
$15$ & $15$ & $1,000$ & $1,000$ & $0.029$ & $0.044$ & $0.042$ & $0.022$ & $0.024$ \\ 
$30$ & $15$ & $1,000$ & $1,000$ & $0.031$ & $0.044$ & $0.042$ & $0.023$ & $0.025$ \\ 
$15$ & $15$ & $2,000$ & $1,000$ & $0.020$ & $0.043$ & $0.041$ & $0.018$ & $0.017$ \\ 
$30$ & $15$ & $2,000$ & $1,000$ & $0.022$ & $0.044$ & $0.041$ & $0.018$ & $0.019$ \\ 
$15$ & $30$ & $500$ & $1,000$ & $0.041$ & $0.044$ & $0.042$ & $0.027$ & $0.030$ \\ 
$30$ & $30$ & $500$ & $1,000$ & $0.043$ & $0.045$ & $0.043$ & $0.028$ & $0.032$ \\ 
$15$ & $30$ & $1,000$ & $1,000$ & $0.029$ & $0.044$ & $0.042$ & $0.023$ & $0.024$ \\ 
$30$ & $30$ & $1,000$ & $1,000$ & $0.031$ & $0.047$ & $0.044$ & $0.024$ & $0.025$ \\ 
$15$ & $30$ & $2,000$ & $1,000$ & $0.020$ & $0.045$ & $0.043$ & $0.018$ & $0.018$ \\ 
$30$ & $30$ & $2,000$ & $1,000$ & $0.022$ & $0.046$ & $0.043$ & $0.018$ & $0.019$ \\ 
$15$ & $15$ & $500$ & $2,000$ & $0.040$ & $0.031$ & $0.030$ & $0.022$ & $0.026$ \\ 
$30$ & $15$ & $500$ & $2,000$ & $0.043$ & $0.032$ & $0.030$ & $0.022$ & $0.027$ \\ 
$15$ & $15$ & $1,000$ & $2,000$ & $0.029$ & $0.031$ & $0.029$ & $0.019$ & $0.022$ \\ 
$30$ & $15$ & $1,000$ & $2,000$ & $0.031$ & $0.031$ & $0.029$ & $0.020$ & $0.022$ \\ 
$15$ & $15$ & $2,000$ & $2,000$ & $0.020$ & $0.032$ & $0.030$ & $0.016$ & $0.016$ \\ 
$30$ & $15$ & $2,000$ & $2,000$ & $0.022$ & $0.032$ & $0.030$ & $0.016$ & $0.017$ \\ 
$15$ & $30$ & $500$ & $2,000$ & $0.040$ & $0.032$ & $0.031$ & $0.022$ & $0.027$ \\ 
$30$ & $30$ & $500$ & $2,000$ & $0.043$ & $0.032$ & $0.030$ & $0.022$ & $0.028$ \\ 
$15$ & $30$ & $1,000$ & $2,000$ & $0.029$ & $0.033$ & $0.030$ & $0.020$ & $0.022$ \\ 
$30$ & $30$ & $1,000$ & $2,000$ & $0.031$ & $0.032$ & $0.030$ & $0.020$ & $0.022$ \\ 
$15$ & $30$ & $2,000$ & $2,000$ & $0.020$ & $0.032$ & $0.030$ & $0.016$ & $0.017$ \\ 
$30$ & $30$ & $2,000$ & $2,000$ & $0.022$ & $0.032$ & $0.030$ & $0.016$ & $0.017$ \\ \hline
\end{tabular} } } 
\end{table} 

\end{document}